\documentclass[12pt,superscriptaddress]{article}

\pdfoutput=1
\usepackage{jheppub}
\usepackage{graphicx}  
\usepackage{dcolumn}   
\usepackage{bm}        
\usepackage{amssymb}   
\usepackage{slashed} 

\newcommand{\beq}{\begin{equation}}
\newcommand{\eeq}[1]{\label{#1}\end{equation}}
\def\beqa{\begin{eqnarray}}
\def\eeqa#1{\label{#1}\end{eqnarray}}
\newcommand{\eeqn}{\end{equation}}
\newcommand{\CR}{\notag \\}
\newcommand{\leqn}[1]{(\ref{#1})}
\def\nn{\nonumber}

\def\stacksymbols #1#2#3#4{\def\theguybelow{#2}
    \def\vp{\lower#3pt}
    \def\sp{\baselineskip0pt\lineskip#4pt}
    \mathrel{\mathpalette\intermediary#1}}

\def\intermediary#1#2{\vp\vbox{\sp
     \everycr={}\tabskip0pt
     \halign{$\mathsurround0pt#1\hfil##\hfil$\crcr#2\crcr
              \theguybelow\crcr}}}

\def\gsim{\stacksymbols{>}{\sim}{2.5}{.2}}
\def\lsim{\stacksymbols{<}{\sim}{2.5}{.2}}
\begin{document}

\title{Spin-One Top Partner: Phenomenology}

\author[a]{Jack H. Collins,} 
\author[b]{Bithika Jain,}
\author[a]{Maxim Perelstein,}
\author[a]{Nicolas Rey-Le Lorier} 

\affiliation[a]{Laboratory for Elementary Particle Physics, 
	     Cornell University, Ithaca, NY 14853, USA}

\affiliation[b]{Department of Physics, Syracuse University, USA}

\emailAdd{mp325@cornell.edu}     

\abstract{Cai, Cheng, and Terning (CCT) suggested a model in which the left-handed top quark is identified with a gaugino of an extended gauge group, and its superpartner is a spin-1 particle. We perform a phenomenological analysis of this model, with a focus on the spin-1 top partner, which we dub the ``swan". We find that precision electroweak fits, together with direct searches for $Z^\prime$ bosons at the LHC, place a lower bound of at least about 4.5 TeV on the swan mass. An even stronger bound, 10 TeV or above, applies in most of the parameter space, mainly due to the fact that the swan is typically predicted to be significantly heavier than the $Z^\prime$. We find that the 125 GeV Higgs can be easily accommodated in this model with non-decoupling D-terms. In spite of the strong lower bound on the swan mass, we find that corrections to Higgs couplings to photons and gluons induced by swan loops are potentially observable at future Higgs factories. We also briefly discuss the prospects for discovering a swan at the proposed 100 TeV $pp$ collider.   
}


\maketitle


\section{Introduction}

Discovery of the Higgs boson brought into sharp focus the long-standing theoretical problem of the Standard Model (SM), the hierarchy problem. If the SM is the complete description of physics up to scale $\Lambda$, radiative corrections generate a contribution to the Higgs mass parameter of order $\Lambda/(4\pi)$. The Higgs mass parameter is now precisely known, $\mu=(126$ GeV$)/\sqrt{2}\approx 90$ GeV. Unless unrelated contributions to $\mu$ cancel, we expect the scale of SM break-down $\Lambda$ to be of order 1 TeV. This argument strongly motivates experimental searches for non-SM physics at the LHC energies, and an extensive program of such searches is ongoing. 

The hierarchy argument does not uniquely fix the nature of new physics at scale $\Lambda$, but it does provide some important clues. Precision electroweak measurements constrain the scale at which generic strong-coupling extensions of the SM may become relevant to $\sim 10$ TeV or above. This indicates that the solution to the hierarchy problem must rely on weakly-coupled physics, unless significant fine-tuning is involved. All known weakly-coupled solutions to the hierarchy problem involve new particles at the scale $\Lambda\lsim$ TeV. Loops of these particles introduce additional contributions to the Higgs mass parameter, which cancel the leading contribution of SM loops. Such cancellations can occur naturally due to symmetries; known examples are supersymmetry, shift symmetry, and gauge symmetry extended to models with extra compact dimensions of space. Each of these symmetries can be implemented in a variety of ways, leading to a large zoo of possible explicit models for non-SM physics at the TeV scale. Most of these models have a rich spectrum of new states, and their masses are typically extremely model-dependent, making it difficult to choose optimal targets for experimental searches. However, in all models, the particles canceling the loops of SM tops, the ``top partners", play a special role. The large value of the top Yukawa in the SM implies that the top partners must be quite light, below a few hundred GeV, for the model to be natural, independent of model-building details. This makes top partners a particularly well-motivated target for the LHC searches.

The conventional wisdom says that top partners fall into one of two classes: spin-0 partners, or ``stops", if the hierarchy problem is solved by supersymmetry; and spin-1/2 partners, if it is solved by shift symmetry or higher-dimensional gauge symmetry. Both these possibilities are extensively covered by experimental searches. There is, however, an alternative possibility, which has so far received far less attention: a spin-1 top partner. An explicit model realizing this scenario was constructed by Cai, Cheng and Terning (CCT) in 2008~\cite{Cai:2008ss}. However, to date, no comprehensive study of phenomenology of this model has been performed. The goal of this paper is to rectify this omission.  

The paper is organized as follows. We review the CCT model, emphasizing the aspects that will be germane for the discussion of phenomenology, in Section~\ref{sec:model}. We then discuss the two main sources of current constraints on the model, precision electroweak fits (Section~\ref{sec:PEW}) and direct searches for $Z^\prime$ bosons at the LHC (Section~\ref{sec:Zprime}). In Section~\ref{sec:HiggsMass}, we discuss how the 125 GeV Higgs boson can be accommodated in this model, and briefly discuss the degree of fine-tuning implied by the constraints. Section~\ref{sec:HC} discusses the deviations in the Higgs couplings to gluons and photons induced by the new particles of the CCT model, while Section~\ref{sec:100tev} contains a brief sketch of the possible signatures of the model at a 100 TeV hadron collider. We conclude in Section~\ref{sec:conc}, and relegate some of the details of the analysis to the Appendix. 

\section{Review of the Model}
\label{sec:model}

The model studied in this paper was proposed by Cai, Cheng and Terning (CCT) in~\cite{Cai:2008ss}. In this section we will review the model.

\subsection{Structure and Particle Content}

	\begin{table}[t]
\centering
\begin{tabular}{l|l*{5}{c}c}
              & $SU(5)$ & $SU(3)$ & $SU(2)$ & $U(1)_H$ & $U(1)_V$  & $U(1)_Y$ \\
\hline
$Q_i$ & $\bold{1}$ & $\square$ & $\square$ & $\frac{1}{6}$ & $0$ & $\frac{1}{6}$  \\
$\overline{u}_i$            & $\bold{1}$ & $\overline{\square}$ & $\bold{1}$ & $-\frac{2}{3}$ &  $0$ & $-\frac{2}{3}$  \\
$\overline{d}_i$           & $\bold{1}$ & $\overline{\square}$ & $\bold{1}$ & $\frac{1}{3}$ &  $0$ & $\frac{1}{3}$ \\
$L_i$    & $\bold{1}$ & $\bold{1}$ & $\square$ & $-\frac{1}{2}$ &  $0$ & $-\frac{1}{2}$  \\
$\overline{e}_i$ & $\bold{1}$ & $\bold{1}$ & $\bold{1}$ & $1$ &  $0$ & $1$  \\
$H$ & $\square$ & $\bold{1}$ & $\bold{1}$ & $\frac{1}{2}$ &  $\frac{1}{10}$ & $\left(\frac{2}{3},\frac{1}{2}\right)$  \\
$\overline{H}$ & $\overline{\square}$ & $\bold{1}$ & $\bold{1}$ & $-\frac{1}{2}$ &  $-\frac{1}{10}$ & $\left(-\frac{2}{3},-\frac{1}{2}\right)$  \\
$\Phi_3$ & $\square$ & $\overline{\square}$ & $\bold{1}$ & $-\frac{1}{6}$ &  $\frac{1}{10}$ & $\left(0,-\frac{1}{6}\right)$  \\
$\Phi_2$ & $\square$ & $\bold{1}$ & $\overline{\square}$ & $0$ &  $\frac{1}{10}$ & $\left(\frac{1}{6},0\right)$  \\
$\overline{\Phi}_3$ & $\overline{\square}$ & $\square$ & $\bold{1}$ & $\frac{1}{6}$ &  $-\frac{1}{10}$ & $\left(0,\frac{1}{6}\right)$  \\
$\overline{\Phi}_2$ &  $\overline{\square}$ & $\bold{1}$ & $\square$ & $0$ &  $-\frac{1}{10}$ & $\left(-\frac{1}{6},0\right)$  \\
\end{tabular}
\label{tab:uvQN}
\caption{Chiral superfields of the model, and their gauge quantum numbers. Here, $i=1\ldots 3$ is the flavor index.}
\end{table}

The CCT model is a supersymmetric gauge theory, based on a gauge group $G=SU(5) \times SU(3) \times SU(2) \times U(1)_H \times U(1)_V$. The matter superfields of the model, and their gauge quantum numbers, are listed in Table~1.
The superpotential has the form
\beqa
W &=& y_1 Q_3 \Phi_3 \overline{\Phi}_2 + \mu_3 \Phi_3 \overline{\Phi}_3 + \mu_2 \Phi_2 \overline{\Phi}_2 + y_2 \overline{u}_3 H \overline{\Phi}_3 + \mu_H H \overline{H} \nn \\ &+& \frac{Y_{Uij}}{M_F} Q_i \overline{u}_j \overline{\Phi}_2 H + \frac{Y_{Dij}}{M_F} Q_i \overline{d}_j \Phi_2 \overline{H} + \frac{Y_{Eij}}{M_F} L_i \overline{e}_j \Phi_2 \overline{H},
\eeqa{SuperPot}
where $i, j=1\ldots 3$ are flavor indices. 
In addition, one must also add soft SUSY-breaking terms generated at some messenger scale $\Lambda$. With the usual motivation of the hierarchy problem, we assume that all soft masses are around the TeV scale; their precise values will not be important for most of our discussion. As will be described in more detail below, SUSY breaking triggers gauge symmetry breaking by causing the four link fields, $\Phi_{2,3}$ and $\overline{\Phi}_{2,3}$, to acquire vacuum expectation values (vevs) of the form
 \begin{align}
\langle \Phi_3 \rangle &=& \left( {\begin{array}{ccccc}
f_3 & 0 & 0 & 0 & 0 \\
0 & f_3 & 0 & 0 & 0  \\
0 & 0 & f_3 & 0 & 0 \\
\end{array} } \right)~, & \qquad & 
\langle \overline{\Phi}_3 \rangle^T &=& \left( {\begin{array}{ccccc}
\overline{f}_3 & 0 & 0 & 0 & 0 \\
0 & \overline{f}_3 & 0 & 0 & 0  \\
0 & 0 & \overline{f}_3 & 0 & 0 \\
\end{array} } \right)~, \nn \\
\langle \Phi_2 \rangle &=& \left( {\begin{array}{ccccc}
0 & 0 & 0 & f_2 & 0 \\
0 & 0 & 0 & 0 & f_2  \\
\end{array} } \right)~, & \qquad & 
\langle \overline{\Phi}_2 \rangle^T &=& \left( {\begin{array}{ccccc}
0 & 0 & 0 & \overline{f}_2 & 0 \\
0 & 0 & 0 & 0 & \overline{f}_2  \\
\end{array} } \right)~.
\label{LinkVEVs}
\end{align}
Given their connection with SUSY breaking, we assume that all $f$'s are at roughly the same scale, $f\sim$ TeV; we will discuss experimental constraints on $f$'s in detail later in this paper. This pattern of vevs breaks $G$ to $G_{\rm SM} = SU(3)_c \times SU(2)_L \times U(1)_Y$, with the $SU(3)_c \times SU(2)_L$ identified with the diagonal linear combination of the $SU(3)\times SU(2)$ subgroup of $SU(5)$, and the additional $SU(3)\times SU(2)$ factor in $G$. The unbroken hypercharge $U(1)_Y$ is given by the linear combination of the diagonal generator $T_{24}$ of $SU(5)$ and the two explicit $U(1)$ factors in $G$: $Y=\frac{1}{\sqrt{15}}T_{24} + H + V$. 
The SM gauge couplings at the scale $f$ are related to the $G$ couplings (denoted by hats):
\beqa
\frac{1}{g_{2,3}^2} = \frac{1}{\hat{g}^2_{2,3}} + \frac{1}{\hat{g}_5^2}~, \qquad \frac{1}{g_{Y}^2} = \frac{1}{\hat{g}^2_{H}} + \frac{1}{\hat{g}_V^2}+\frac{1}{15\hat{g}_5^2}.
\eeqa{GaugeRel}
Examining the matter field quantum numbers under $G_{\rm SM}$, it is easily seen that the model contains all of the familiar matter content of the MSSM. In particular, the fields $Q_i$, $\bar{u}_i$, 
$\bar{d}_i$, $L_i$ and $\bar{e}_i$ are directly identified with the corresponding MSSM fields, with the exception of the third-generation quarks which require special treatment. The two Higgs fields of the MSSM, $H_d$ and $H_u$, are embedded in the $H$ and $\overline{H}$ fields, along with the (non-MSSM) color triplets and anti-triplets $\overline{T}^c$ and  
$\overline{T}$:
\beqa
H = \left( {\begin{array}{c}
\overline{T}^c \\
H_u \\
\end{array} } \right)~, \qquad
\overline{H} = \left( {\begin{array}{c}
\overline{T} \\
H_d \\
\end{array} } \right)~.
\eeqa{Hfields}
The last four terms of the superpotential~\leqn{SuperPot} then reproduce the full MSSM superpotential. In particular, SM quark and lepton Yukawa couplings are of order $f/M_F$, and can naturally be small if there is a hierarchy between these scales. 

 \begin{table}[t]
 \begin{center}
 \small{
\begin{tabular}{ |l|l|l|l|l|l|l|l| }
\hline
Field & Spin & $SU(3)_c$ & $SU(2)_L$ & $U(1)_Y$ & R-Parity &UV Multiplet  &Mass \\ 
& & & & & & & Scale \\ \hline
$\Phi_{3S}$, $\overline{\Phi}_{3S}$ & 0 & $\bold{1}$ & $\bold{1}$ & 0 & +1&$\Phi_i$, $\overline{\Phi}_i$& $f$ \\
$\Phi_{2S}$, $\overline{\Phi}_{2S}$& & & & & &  &\\ \hline
$\Phi_{3A}$, $\overline{\Phi}_{3A}$ & 0 & $\bold{Adj}$ & $\bold{1}$ & 0 & +1 & $\Phi_3$, $\overline{\Phi}_3$&$f$ \\ \hline
$\Phi_{2A}$, $\overline{\Phi}_{2A}$ & 0 & $\bold{1}$ & $\bold{Adj}$ & 0 & +1 & $\Phi_2$, $\overline{\Phi}_2$&$f$ \\ \hline
$\widetilde{\overline{\Phi}}_{3t}$, $\widetilde{\Phi}_{2t}$ & 0 & $\bold{3}$ & $\bold{2}$ & 1/6 & -1& $\Phi_2$, $\overline{\Phi}_3$&$f$ \\ \hline
$\widetilde{\Phi}_{3t}$, $\widetilde{\overline{\Phi}}_{2t}$ & 0 &$\overline{\bold{3}}$ &$\bold{2}$ &-1/6 & -1&$\Phi_3$, $\overline{\Phi}_2$ & $f$\\ \hline
$\widetilde{\Phi}_{3S}$, $\widetilde{\overline{\Phi}}_{3S}$ & 1/2 & $\bold{1}$ & $\bold{1}$ & 0 & -1&$\Phi_i$, $\overline{\Phi}_i$& $f$ \\
$\widetilde{\Phi}_{2S}$, $\widetilde{\overline{\Phi}}_{2S}$& & & & & &  &\\ \hline
$\widetilde{\Phi}_{3A}$, $\widetilde{\overline{\Phi}}_{3A}$ & 1/2 & $\bold{Adj}$ & $\bold{1}$ & 0 & -1 & $\Phi_3$, $\overline{\Phi}_3$&$f$ \\ \hline
$\widetilde{\Phi}_{2A}$, $\widetilde{\overline{\Phi}}_{2A}$ & 1/2 & $\bold{1}$ & $\bold{Adj}$ & 0 & -1 & $\Phi_2$, $\overline{\Phi}_2$&$f$ \\ \hline
$\overline{\Phi}_{3t}$, $\Phi_{2t}$ & 1/2 & $\bold{3}$ & $\bold{2}$ & 1/6 & +1& $\Phi_2$, $\overline{\Phi}_3$&$f$ \\ \hline
$\Phi_{3t}$, $\overline{\Phi}_{2t}$ & 1/2 &$\overline{\bold{3}}$ &$\bold{2}$ &-1/6 & +1&$\Phi_3$, $\overline{\Phi}_2$ & $f$\\ \hline
$\widetilde{\overline{T}}$ & 0 & $\overline{\bold{3}}$ & $\bold{1}$ & -2/3 & -1 & $\overline{H}$ & $f$ \\ \hline
$\widetilde{\overline{T}^c}$ & 0 & $\bold{3}$ & $\bold{1}$ & 2/3 & -1 & $H$ & $f$ \\ \hline
$\overline{T}$ & 1/2 & $\overline{\bold{3}}$ & $\bold{1}$ & -2/3 & +1 & $\overline{H}$ & $v$ \\ \hline
$\overline{T}^c$ & 1/2 & $\bold{3}$ & $\bold{1}$ & 2/3 & +1 & $H$ & $f$ \\ \hline
$\lambda$ & 1/2 & $\bold{3}$ & $\bold{2}$ & 1/6 & +1 & $SU(5)$ gauginos&$v$\\ \hline
$\overline{\lambda}$ & 1/2 & $\overline{\bold{3}}$ & $\bold{2}$ & -1/6 & +1 & $SU(5)$ gauginos&$f$\\ \hline
$\widetilde{W}'$ & 1/2 & $\bold{1}$ & $\bold{Adj}$ & 0 & -1 & $SU(2)$, $SU(5)$ gauginos& $f$ \\ \hline
$\widetilde{G}'$ & 1/2 & $\bold{Adj}$ & $\bold{1}$ & 0 & -1 &  $SU(3)$, $SU(5)$ gauginos& $f$ \\ \hline
$\widetilde{B}'$, $\widetilde{B}''$ & 1/2 & $\bold{1}$ & $\bold{1}$ & 0 & -1 &  $U(1)_H$, $U(1)_V$, $SU(5)$ gauginos& $f$ \\ \hline
$W'$ & 1 & $\bold{1}$ & $\bold{Adj}$ & 0 & +1 & $SU(2)$, $SU(5)$ gauge fields& $f$ \\ \hline
$G'$ & 1 & $\bold{Adj}$ & $\bold{1}$ & 0 & +1 &  $SU(3)$, $SU(5)$ gauge fields& $f$ \\ \hline
$Z'$, $Z''$ & 1 & $\bold{1}$ & $\bold{1}$ & 0 & +1 &  $U(1)_H$, $U(1)_V$, $SU(5)$ gauge fields& $f$ \\ \hline
$\vec{Q}$ & 1 & $\bold{3}$ & $\bold{2}$ & 1/6 & -1 & $SU(5)$ gauge fields& $f$ \\ \hline
\end{tabular}
}
\end{center}
\label{tab:FieldContentSM}
\caption{Field content after the UV symmetry breaking; all entries with spin 0 correspond to complex scalar fields. The MSSM fields are not included in this table.}
\end{table}

\begin{table}[t]
\centering
\begin{tabular}{l|c c c c}
              & $\lambda$ & $\Phi_{2t}$ & $\overline{\Phi_{3t}}$ & $Q_3$ \\
\hline
$\overline{\lambda}$ & $M_5$ & $\hat{g}_5 f_2$ & $\hat{g}_5 \overline{f}_3$ & $0$ \\
$\Phi_{3t}$ & $\hat{g}_5 f_3$ & $0$ & $\mu_3$ & $y_1 \overline{f}_2$ \\
$\overline{\Phi}_{2t}$ & $\hat{g}_5 \overline{f}_2$ & $\mu_2$ & $0$ & $y_1 f_3$ \\
\end{tabular}
\caption{Mass matrix for fermions in the $\left( \bold{3},\bold{2},1/6\right)$ (and conjugate) sector.}
\label{tab:Qmass}
\end{table}

The model also has a rich spectrum of non-MSSM fields. These are listed in Table~2
, along with their $G_{\rm SM}$ quantum numbers and $R$ parity. Since SUSY breaking and $G\to G_{\rm SM}$ breaking occur at roughly the same scale, in this case we list each field and its superpartner separately. Note that the conserved $R$ parity in the CCT model, which plays the same role as the usual $R$ parity in the MSSM, is a convolution of a ``global" $R$ parity which commutes with all gauge transformations, and a ``twist" transformation, which acts on the $SU(5)$ multiplets as $P_{\rm twist}=$~diag$(-1,-1,-1,1,1)$. The twist is required because the scalar components of the $H$ and $\bar{H}$ multiplets must be assigned opposite $R$-parities, $+1$ for the Higgs and $-1$ for the $\overline{T}$ and $\overline{T}^c$. 

Interestingly, some of the fields in Table~2 
have the same quantum numbers as MSSM fields, allowing them to mix. In particular, there are three fields with the quantum numbers of the left-handed quark doublet $Q$, $({\bf 3}, {\bf 2}, 1/6)$: the ``off-diagonal" $SU(5)$ gaugino $\lambda$,  
and the link field ``inos" $\Phi_{2t}$ and $\bar{\Phi}_{3t}$. There are also three fields in the conjugate representation, $(\bar{{\bf 3}}, {\bf 2}, -1/6)$: $\bar{\lambda}$, $\bar{\Phi_{3t}}$ and $\Phi_{2t}$. The mass matrix for these fields, before electroweak symmetry breaking (EWSB), is given in Table~\ref{tab:Qmass}. Note that only $Q_3$ participates in the mixing due to the structure of the superpotential; more generally, we can always relabel the linear combination of the quark doublet fields which couples to $\Phi_3\bar{\Phi}_2$ as $Q_3$. Because the mass matrix has four columns but only three rows, there will always be a linear combination of $Q$-like fields which will be massless at this level, acquiring a mass through ESWB only. We identify this field with the third generation quark doublet of the SM, $Q_3^{SM}$. The key idea of the CCT model is that for a certain range of parameters, $Q_3^{SM}$ is predominantly the gaugino $\lambda$. If that's the case, top-loop contribution to the Higgs mass must be canceled by its superpartner, a spin-1 (``swan") gauge boson $\vec{Q}$. This occurs if~\cite{Cai:2008ss} \beqa
M_5 &\ll& \hat{g}_5 f_2,~~~
\hat{g}_5 f_3 \ll \mu_3,~~~
\hat{g}_5 f_3 \ll \hat{g}_5 \overline{f}_2, \CR
\hat{g}_5 \overline{f}_2 &\ll& \mu_2,~~~
\hat{g}_5^2 \frac{f_2 \overline{f}_2}{M_5 \mu_2} \approx 1,~~~
\hat{g}_5 \lesssim y_1.
\eeqa{LambdaTopConditions}
We will assume throughout this paper that these conditions are realized. Another sector in which mixing occurs is the fields with the quantum numbers $(\bar{{\bf 3}}, {\bf 1}, -2/3)$: $\bar{u}$ and $\bar{T}$. One of their linear combinations gets a mass of order $f$, while the other remains massless until EWSB, and is identified with the SM right-handed top. Generating an order-one top Yukawa requires that the massless combination be predominantly $\overline{T}$; the condition for this is
\beq
\mu_H \ll y_2 \overline{f}_3.
\eeq{TbTopCondition}
The dominant coupling of the SM top to the Higgs comes from the $SU(5)$ gaugino-sfermion-fermion interaction of the field $\overline{H}$:
 \beq
-\sqrt{2}\hat{g}_5\overline{H}^{*}\left(-T^{a*}  \lambda_5^a\right)\widetilde{\overline{H}}  \supset \hat{g}_5 H_d^{*}\lambda \overline{T}.
\eeq{ThirdGenYukawa}
Since $\hat{g}_5$ can be $O(1)$ while the other Yukawa couplings in Eq.~\leqn{SuperPot} are suppressed by the ratio $f/M_F$, this explains the mass splitting between the top and the other quarks. The down-type third generation singlet is still $\overline{d}_3$, just like in the MSSM, so the bottom quark still gets its mass from the superpotential Yukawas. From now on we will assume that the gaugino fraction of the third generation doublet is very close to unity, i.e. $\langle Q_3^{SM} | \lambda \rangle \approx 1$. Note that this equality cannot be exact without forcing the bottom quark's mass to vanish since it is proportional to $\left| \langle Q_3^{SM} | Q_3 \rangle \right| \le \sqrt{1 - \left|\langle Q_3^{SM} | \lambda \rangle \right|^2}$. Still, assuming that the deviation of $\langle Q_3^{SM} | \lambda \rangle$ from unity is small, the gauge coupling $\hat{g}_5$ must satisfy
 \beqa
\hat{g}_5 = \frac{\sqrt{2}m_t}{v\cos{\beta}} \approx \sqrt{1+\tan^2{\beta}}~,
\eeqa{g5}
where $m_t$ is the top mass, $v= \sqrt{v_u^2 + v_d^2} = 246$ GeV, and $\beta$ is defined through the usual MSSM relationship $\tan{\beta} \equiv v_u/v_d$. With this result, the first of Eqs.~\leqn{GaugeRel} uniquely fixes $\hat{g}_2$ and $\hat{g}_3$ in terms of $\tan\beta$, while the second one reduces to
\beq
\frac{1}{g_{Y}^2}(1-\epsilon) = \frac{1}{\hat{g}^2_{H}} + \frac{1}{\hat{g}_V^2},
\eeq{buji}
where 
\beq
\epsilon = \frac{g_Y^2}{15\hat{g}_5^2} \approx \frac{8\cdot 10^{-3}}{1+\tan^2\beta}.
\eeq{epsdef}
Thus, requiring that the model reproduce the SM gauge couplings and the top Yukawa leaves only two independent parameters in the gauge sector: $\tan\beta$ and the $U(1)$-sector mixing angle
\beq
\theta = \arctan \left(\frac{\hat{g}_V}{\hat{g}_H}\right).
\eeq{thetadef}

\subsection{Gauge Boson Spectrum}

\begin{figure}[tb]
\begin{center}
\centerline {
\includegraphics[width=3.0in]{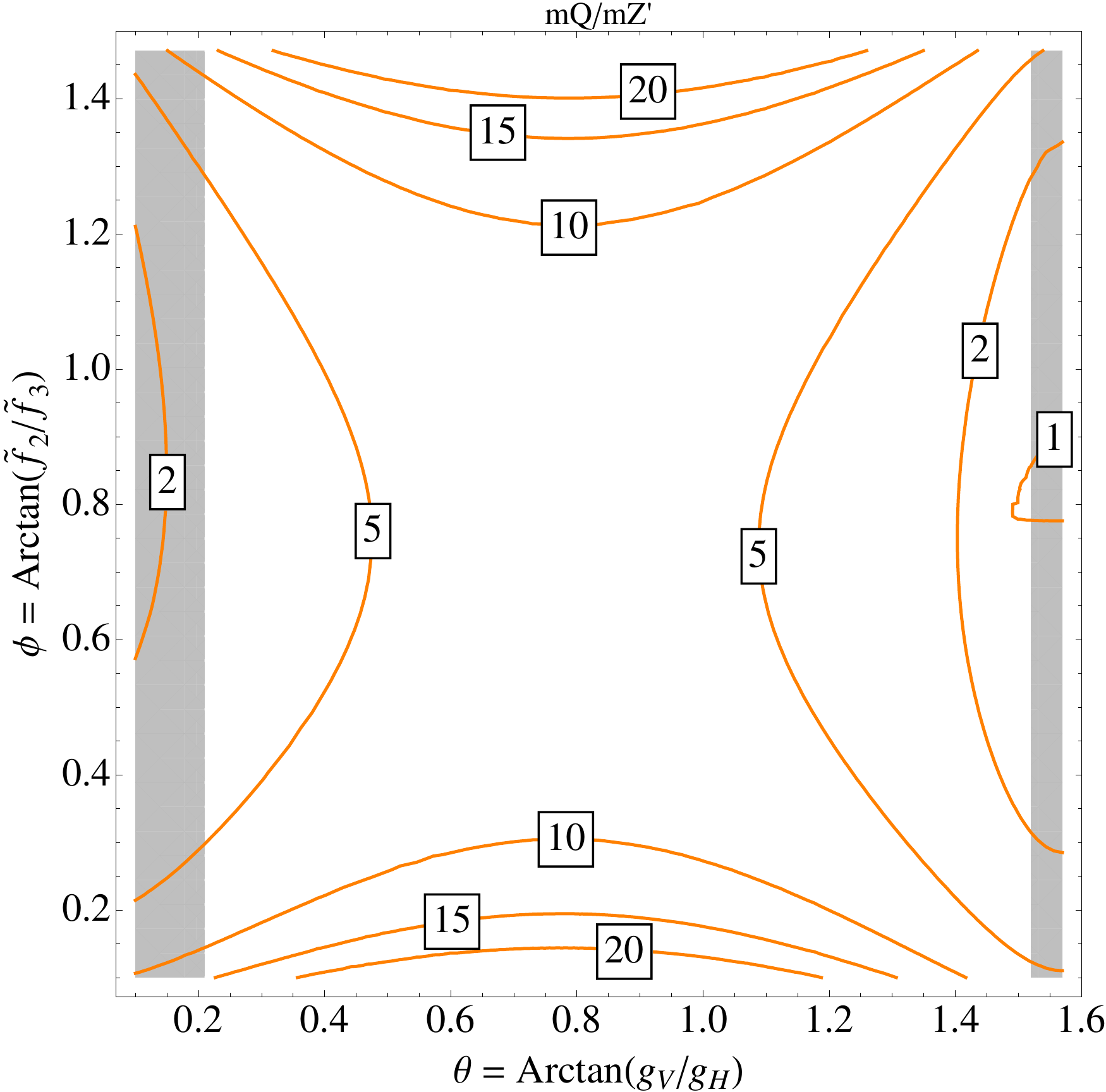}
\includegraphics[width=3.0in]{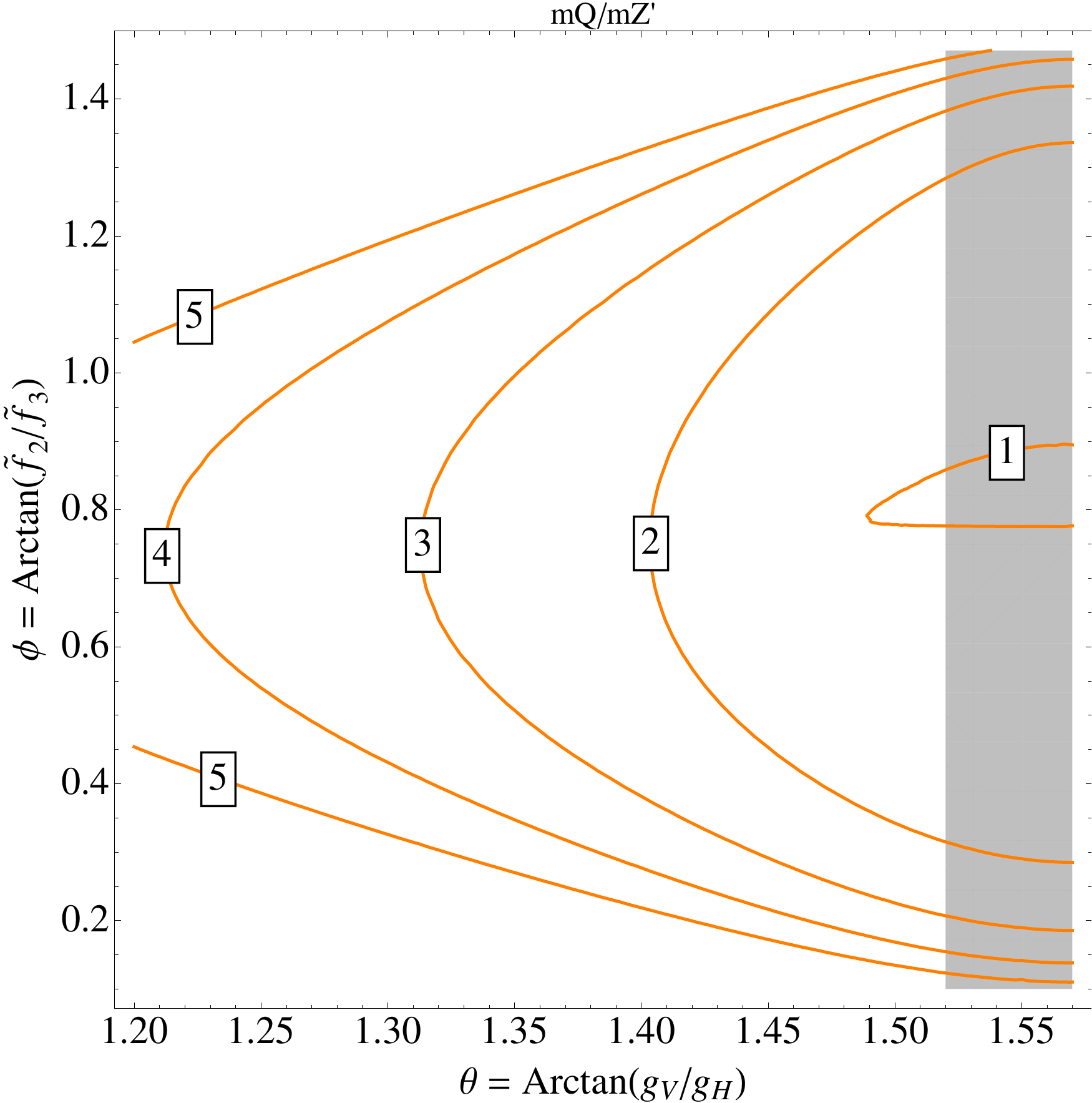}
}
\caption{Ratio of the masses of the spin-1 top partner (``swan") and the lightest $Z^\prime$. Left panel: full parameter space (gray regions indicate regions where one of the gauge couplings becomes non-perturbative). Right panel: the region where the ratio is minimized. In both plots, $\tan\beta=0.95$; the ratio scales as $\sqrt{1+\tan^2\beta}$.}
\label{fig:MQvsMZ}
\end{center}
\end{figure}

The model contains several additional gauge bosons, which will be especially important for the analysis of this paper for two reasons. First, as already mentioned, one of them, the swan $\vec{Q}$, is largely responsible for canceling the quadratically divergent contribution of the SM top loop to the Higgs mass. 
Second, the extra $U(1)$ gauge bosons are responsible for the strongest experimental constraints on the model parameter space. The swan mass is given by
\beq
m^2_{\vec{Q}} = \hat{g}_5^2  \left(\tilde{f}_2^2 + \tilde{f}_3^2 \right),
\eeq{SwanMass}
where we defined 
\beq
\tilde{f}_{2,3} = \frac{f_{2,3}^2+\bar{f}_{2,3}^2}{2}.
\eeq{ftildedefs}
Requiring that the left-handed top quark is predominantly a gaugino requires $f_3 \ll \overline{f}_2$, as mentioned above; however, no particular hierarchy between $\overline{f}_3$ and $f_2$ is required, so the scales $\tilde{f}_2$ and $\tilde{f}_3$ are essentially independent parameters. We find it convenient to define
\beq
\tilde{f} = \sqrt{\tilde{f}_2^2 +  \tilde{f}_3^2},~~~\phi = \arctan \frac{\tilde{f}_2}{\tilde{f}_3}. 
\eeq{phidef}
With this notation, the swan mass is simply 
\beq
m^2_{\vec{Q}} = \hat{g}_5^2 \tilde{f}^2 \approx (1+\tan^2\beta) \tilde{f}^2.
\eeq{swanmass2} 
The mass of the lightest extra $U(1)$ gauge boson, the $Z^\prime$, is given by
\beq
m_{Z^\prime}^2 \approx 2 g_Y^2 \frac{\csc^2 2\theta \sin^2 2\phi}{5 - \cos 2\phi}\,\tilde{f}^2,
\eeq{Zpmassapprox}
where corrections of order $\epsilon$ and $v^2/\tilde{f}^2$ have been dropped. (The complete spectrum of the $U(1)$ gauge bosons is given in Appendix A.) Since $g_Y\approx 0.3$, the swan is generally significantly heavier that the $Z^\prime$; see Fig.~\ref{fig:MQvsMZ}. We will see below that this results in very strong experimental lower bounds on the swan mass. 

For completeness, we also list the masses of the heavy partners of the gluon and the charged $W$ bosons: 
\beqa
m^2_{G'} &=& 2 \left(\hat{g}_3^2 + \hat{g}_5^2 \right) \tilde{f}_3^2 \approx \frac{2g_3^2 (1+\tan^2\beta) \cos^2\phi}{1+\tan^2\beta-g_3^3}\,\tilde{f}^2, \\
m^2_{W'} &=& 2 \left(\hat{g}_2^2 + \hat{g}_5^2 \right) \tilde{f}_2^2 \approx \frac{2g_2^2 (1+\tan^2\beta) \sin^2\phi}{1+\tan^2\beta-g_2^3}\,\tilde{f}^2.
\eeqa{GaugeFieldMasses}

\subsection{Beta Functions and the Strong-Coupling Scale}

The CCT model is an effective theory, since some of its gauge groups are not asymptotically free and their gauge couplings hit a Landau pole at a finite energy scale. At that scale, the model has to be either embedded into a larger structure, providing a UV completion, or else a non-perturbative description of the dynamics is required. Defining the one-loop beta function as
\beq
\beta_i \equiv \mu \frac{dg_i}{d\mu}=- \frac{g_i^3}{16\pi^2}b_i,
\eeq{betadef}
we find the coefficients
\beq
b_5 = 9 ~,
b_3 = -2~,
b_2 = -5~,
b_H = - \frac{40}{3}~,
b_V = - \frac{3}{5}~.
\eeq{bs}
With the exception of $SU(5)$, all other factors in $G$ are not asymptotically free. We estimate the strong-coupling scale $\Lambda_i$ for each group with the condition $g_i(\Lambda_i) = \beta_i$, or equivalently $b_ig_i^2/(16\pi^2) = 1$; this yields 
\beq
\Lambda_i = f_i \exp\left[ \frac{2 \pi}{|b_i| \alpha_i(f)} - \frac{1}{2}\right],
\eeq{Lambdas}
where $f_i$ is the scale where the gauge group associated with each gauge coupling is broken. 

The parameters in the gauge sector of the theory are restricted by perturbativity requirements. For the asymptotically free $SU(5)$ coupling, we demand $b_5 \hat{g}_5^2/(16\pi^2) \leq 1$ at the symmetry-breaking scale $f$; for the other couplings, we require $\Lambda_i/f \gsim 5$. This yields
\beqa
0.8 \lsim \tan{\beta} \lsim 4.0,~~0.2 \lsim \sin\theta \lsim 0.99.
\eeqa{TanBetaConstraints}
The bounds on $\tan\beta$ should be compared to the case of the MSSM, where the relationship analogous to Eq.~(\ref{g5}) is $y_t = \frac{\sqrt{2}m_t}{v\sin{\beta}}$ and imposes only the much weaker constraints $0.3 \lsim \tan{\beta} \lsim 150$. The fact that $\tan{\beta}$ is constrained to lie close to $1$ will tend to suppress the Higgs mass, since at tree-level and in the decoupling limit it is proportional to $\cos2\beta$; this will be discussed in Section~\ref{sec:HiggsMass}.

\section{Precision Electroweak Constraints}
\label{sec:PEW}

As described in the previous section, the CCT model extends the SM gauge group and introduces additional R-even gauge bosons, $W^\prime$ and $Z^\prime$. These gauge bosons generically mix with the SM $Z$ and $W$, leading to deviations of their properties from the SM predictions. In addition, tree-level exchanges of  $W^\prime$ and $Z^\prime$ induce effective four-fermion interactions not present in the SM. Such effects are tightly constrained by precision electroweak (PEW) measurements, which can be translated into restrictions on the parameter space of the CCT model. Before proceeding, let us note that while the CCT model predicts many new states at the TeV scale (see Table~\ref{tab:FieldContentSM}), it is easy to see that the PEW constraints are dominated by the $W^\prime$ and $Z^\prime$. Most of the other fields do not contribute to PEW observables at tree level at all, either due to negative $R$-parity or, as in the case of vector-like fermions in the top sector and the heavy partner of the gluon, due to the structure of their couplings to the SM. The only states that do make a tree-level contribution are the scalars from link fields, which however only have suppressed couplings to light fermions of order $v/M_F$. We will ignore such contributions. 

\begin{figure}[tb]
\begin{center}
\centerline {
\includegraphics[width=3.0in]{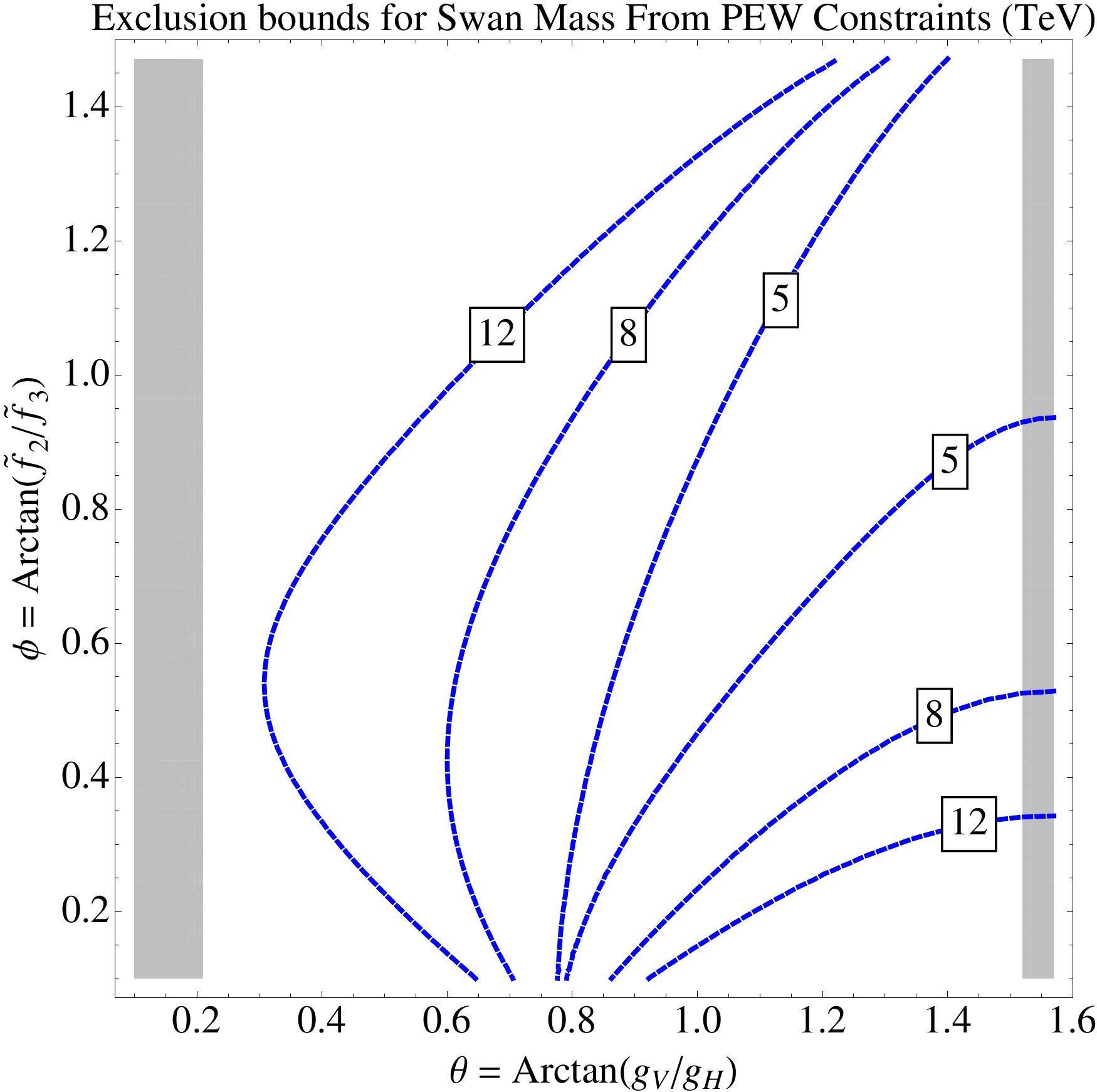}
\includegraphics[width=3.0in]{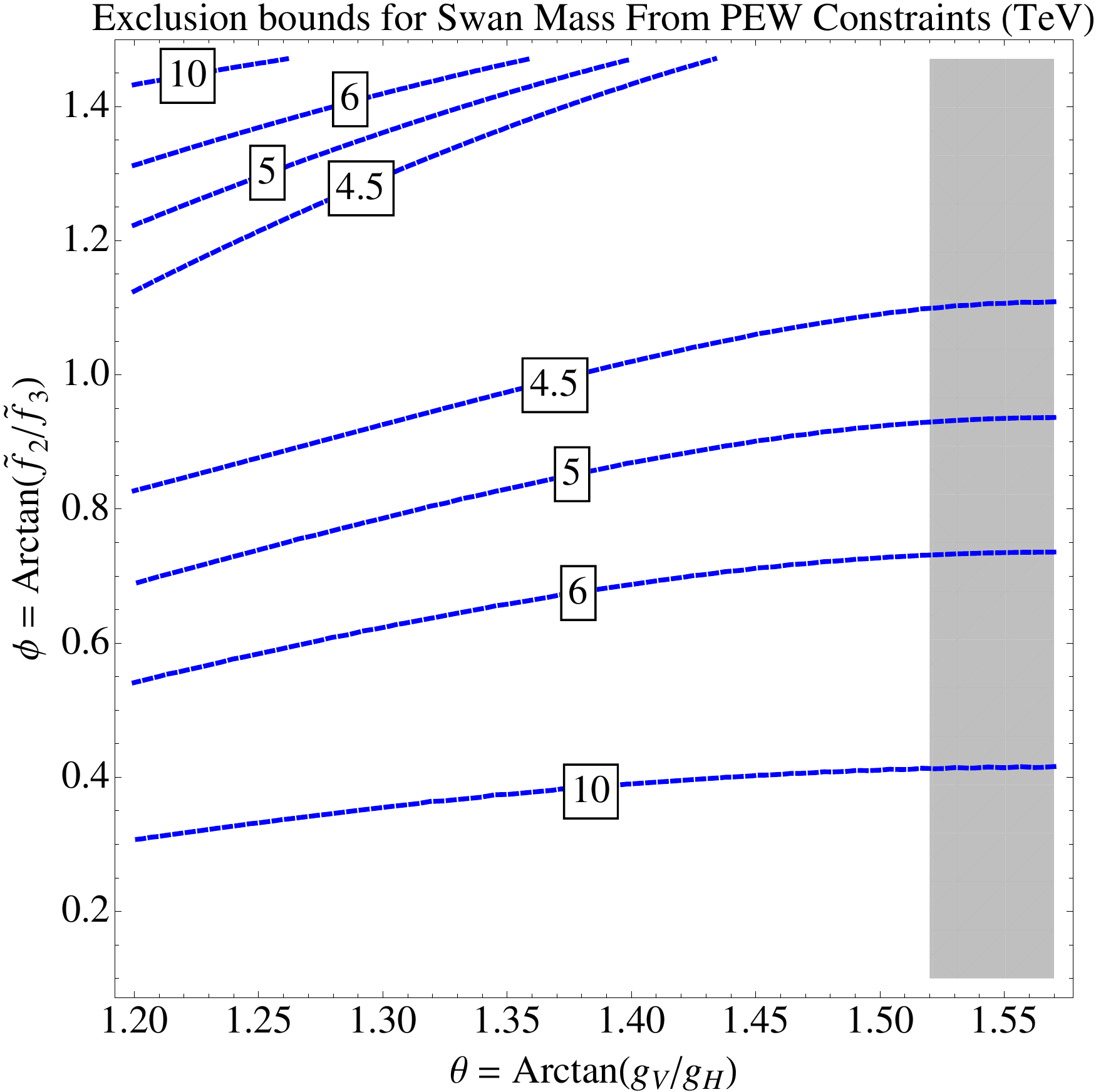}
}
\caption{Lower bound on the swan mass (in TeV) from precision electroweak constraints. Left panel: full parameter space (gray regions indicate regions where one of the gauge couplings becomes non-perturbative). Right panel: the region where the constraint is minimized. In both plots, $\tan\beta=0.95$; the bounds scale as $\sqrt{1+\tan^2\beta}$.}
\label{fig:PEWs1}
\end{center}
\end{figure}

It is well known that the effect of $Z^\prime$ and $W^\prime$ bosons on PEW observables can be cast in terms of the oblique parameters $S$, $T$ and $U$~\cite{Langacker:1991pg,Holdom:1990xp,Altarelli:1991fk,Barbieri:2004qk}. Evaluating the $T$ parameter in the CCT model yields\footnote{Oblique parameters in the CCT model have been previously computed in Ref.~\cite{Cai:2012bsa}.} 
\beq
\alpha T  = \left[ \frac{3}{4}\frac{\left( (1-\epsilon) \cos^2\theta + \epsilon \right)^2}{\cos^2\phi}  + \frac{1}{8} \frac{\left((1-\epsilon) \cos2\theta - 4 \epsilon \right)^2}{\sin^2\phi}\right] \, 
\left(\frac{v}{\tilde{f}}\right)^2.
\eeq{Tpar}
Both $S$ and $U$ parameters are not generated at order $(v/\tilde{f})^2$. The leading contributions to these parameters, up to ${\cal O}(\epsilon)$ corrections, are given by 
\beqa
U &=&  \left( \frac{\cos^2{\theta_W}}{2\alpha} \right)\left(\frac{9\sin^2{\theta}\cos^6{\theta}}{2\cos^4{\phi}}  + \frac{\sin^2{4\theta}}{32 \sin^4{\phi}} + \frac{3 \sin^2{\theta}\cos{2\theta}\cos^4{\theta}}{\sin^2{\phi} \cos^2{\phi}} \right) \left(\frac{v}{\tilde{f}}\right)^4; \CR
S &=& -U - \frac{\sin^2\theta_W}{16\alpha}\,\frac{1}{\sin^4\phi}\,x\left(1+x\right)^{-3}\,\left(\frac{v}{\tilde{f}}\right)^4,
\eeqa{SUpars}
where $x=(g_2/\hat{g}_5)^2 \approx g_2^2 (1+\tan^2\beta)^{-1}$. 

The 95\%~c.l. PEW lower bound on the swan mass is shown in Figure~\ref{fig:PEWs1}. As expected, the bound is strongly dominated by the $T$ parameter. (The current 95\%~c.l. bound on $T$, for $S\approx0$, is $T\lsim 0.12$~\cite{Baak:2013ppa}.) Here we fixed $\tan\beta=0.95$, close to the low end of the allowed range; the bound is stronger for larger values of $\tan\beta$, scaling as $\sqrt{1+\tan^2\beta}$. We find that the lowest possible bound occurs when $\tilde{f}_2 > \tilde{f}_3$ and $\hat{g}_V \gg \hat{g}_H$, and it is roughly given by
\beq
m_{\vec{Q}} \gsim 4.5~{\rm TeV}.
\eeq{PEWbound}
Since swans need to be pair-produced in proton collisions due to their negative R-parity, this bound effectively puts them out of reach of the direct LHC searches. It also implies significant fine-tuning in the EWSB, as will be discussed in section~\ref{sec:HiggsMass}. 

Additional contributions to PEW observables may be generated by strongly-coupled physics in the ultraviolet (UV), and in a generic UV completion, the strong-coupling scale must be above $\sim10$ TeV to avoid conflict with experiment. Bounds on the perturbative contribution to the $T$ parameter, together with the parameter space constraints~\leqn{TanBetaConstraints}, ensure that such non-perturbative contributions are negligible throughout the viable parameter space, with the possible exception of the far upper-right corner of the plots in Fig.~\ref{fig:PEWs1}, where the $SU(3)$ gauge group may become strongly coupled below 10 TeV. Since $SU(3)$ is not part of the electroweak gauge group, this by itself does not imply additional contributions to PEW observables at the same scale; they may or may not be induced, depending on the nature of the UV completion. In any case, this caveat only affects a small corner of the parameter space, and the basic conclusions of the perturbative analysis remain valid.


\section{Direct Searches at the LHC}
\label{sec:Zprime}

Further bounds on the model parameter space come from direct searches at the LHC. Conventional SUSY searches place bounds on many of the R-odd states, which are also present in the MSSM spectrum. In the MSSM, assuming a spectrum with a weakly interacting lightest R-odd particle, and large mass gaps between this particle and colored R-odd states, current LHC bounds require $m_{\tilde{G}}\gsim 1.2-1.4$ TeV for gluinos, $m_{\tilde{Q}}\gsim 0.8$ TeV for squarks of first two generations, and $m_{\tilde{t}}\gsim 0.7$ TeV for stops/sbottoms. The bounds in the CCT model can be modified due to the presence of additional states with the quantum numbers of gluinos and stops, $\tilde{G}^\prime$, $\tilde{\bar{T}}$, and $\tilde{\bar{T}}^\prime$. These can induce additional cascade decays, strengthening the bounds somewhat; however, we do not expect a major qualitative change. It should also be noted that while the superpartner masses are generally expected to be at the scale $f$, the precise relation between them is model-dependent, since the details of SUSY breaking come into play. On the other hand, searches for the R-even states, in particular extra gauge bosons, in many cases have higher reach, since these states can be produced singly, and can be described in terms of just a small number of parameters, as explained in Section~\ref{sec:model}. With this motivation, we investigate these bounds in detail in this section.

\begin{figure}[tb]
\begin{center}
\centerline {
\includegraphics[width=3.0in]{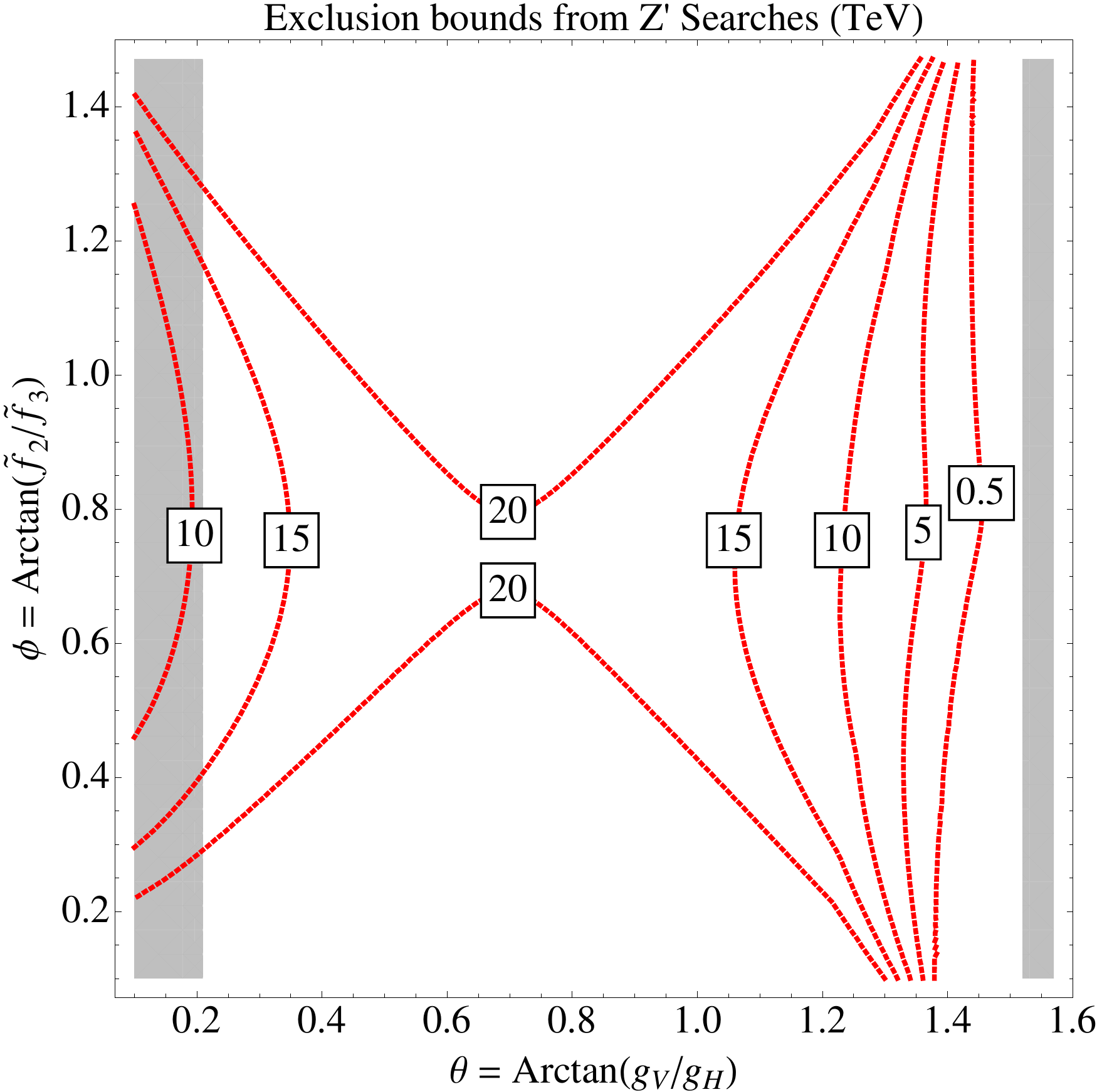}
\includegraphics[width=3.0in]{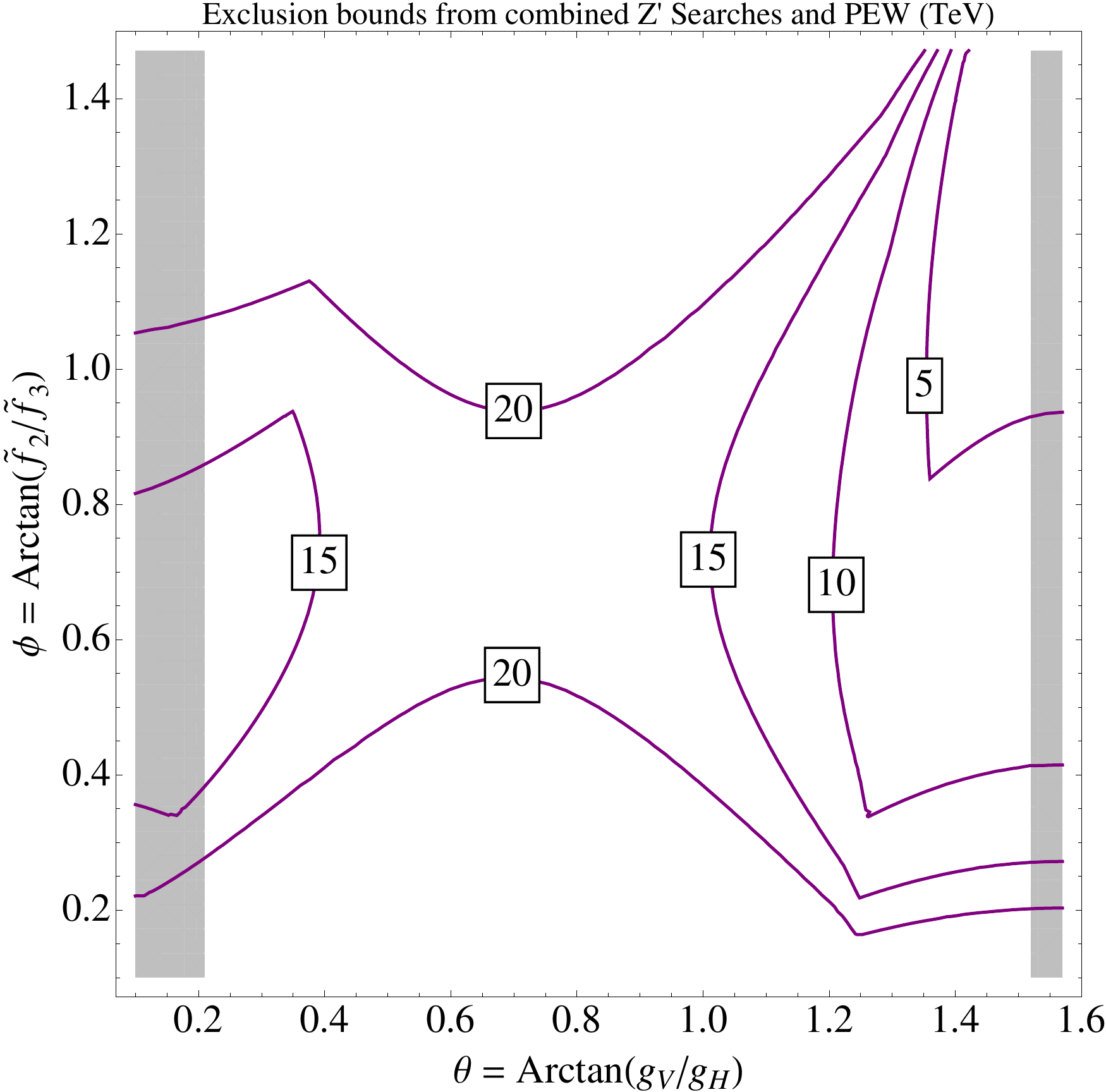}
}
\caption{Lower bounds on the swan mass (in TeV) from direct searches for the $Z^\prime$ at the LHC (left panel) and the combination of direct search and precision electroweak constraint (right panel). In both plots, $\tan\beta=0.95$; the bounds scale as $\sqrt{1+\tan^2\beta}$.}
\label{fig:Direct}
\end{center}
\end{figure}

The strongest bounds come from searches for $Z^\prime$ gauge bosons, in particular in the $Z^\prime \to \mu^+\mu^-$ channel. We incorporated the relevant couplings of the CCT model (listed in Appendix~\ref{app:Zprime}) into the {\tt MadGraph/MadEvent 5} event generator~\cite{Alwall:2011uj}, and computed the cross section of the process $pp\to Z^\prime \to\mu^+\mu^-$ at the $\sqrt{s}=8$ TeV LHC as a function of the $Z^\prime$ mass. We then used the cross section bound presented by the CMS collaboration~\cite{CMS-PAS-EXO-12-061}, based on the full 20 fb$^{-1}$ data set collected at LHC-8, to constrain the model parameter space. The resulting bound on the swan mass, for $\tan\beta=0.95$, is shown in Fig.~\ref{fig:Direct} (left panel). 
(As for precision electroweak, the direct search bound on the swan mass scales as $\sqrt{1+\tan^2\beta}$, so the bounds in Fig.~\ref{fig:Direct} become stronger for larger $\tan\beta$.) 
Generically, the bounds on the swan mass are quite high, above 10 TeV in most of the parameter space. This is stronger than the PEW bound. However, the direct search bound is weakened significantly in the region $g_V\gg g_H$, where the $Z^\prime$ couplings to fermions are suppressed. In this region, the PEW constraint dominates; the combined bound from PEW and direct searches is presented in Fig.~\ref{fig:Direct} (right panel). Overall, the lowest bound on $m_{\vec{Q}}$ found in the PEW analysis, about 4.5 TeV, remains unchanged.  

In addition to $Z^\prime$, the model contains two more electrically neutral gauge bosons: $Z^{\prime\prime}$, the heaviest of the mass eigenstates composed of $U(1)_H$, $U(1)_V$ and $T_{24}$ gauge bosons; and $W^{\prime 3}$, the heavy mass eigenstate composed of the diagonal $SU(2)$ and $SU(2)^\prime\in SU(5)$ gauge bosons. 
Since $\hat{g}_5$ is larger than the other gauge couplings, both $Z^{\prime\prime}$ and $W^{\prime 3}$ are significantly heavier than the $Z^\prime$ throughout the parameter space. Furthermore, for the same reason, both $Z^{\prime\prime}$ and $W^{\prime 3}$ are dominated by their $SU(5)$ components, and since light fermions are not charged under the $SU(5)$, their production cross sections are suppressed. As a result, we find that including these states in the analysis does not improve the bounds derived by considering only the lightest $Z^\prime$. Likewise, massive electrically charged gauge bosons $W^\prime$ and color-octet gauge bosons $G^\prime$ do not yield relevant bounds. 

\section{Higgs Mass and EWSB Fine-Tuning}
\label{sec:HiggsMass}

Just as in the MSSM, the superpotential of the CCT model, Eq.~\leqn{SuperPot}, does not contribute to the Higgs quartic coupling, and the D-term contribution by itself is far too small for compatibility with a 125 GeV Higgs. The quartic is enhanced by the RG evolution between the SUSY breaking scale $\Lambda_{\rm susy}$, and the electroweak scale. To understand whether this enhancement is sufficient to produce a viable Higgs mass, we evolve the weak-scale Higgs quartic $\lambda(M_t)$, inferred from the data, up to the scale $\Lambda_{\rm susy}$, and compare it with the SUSY prediction at that scale:\footnote{Our normalization for $\lambda$ is such that the Higgs scalar potential in $V=-m^2H^\dagger H+\lambda (H^\dagger H)^2$, where $H$ is the Higgs doublet field. In this normalization, $\lambda_{\rm SM}(M_t) \approx 0.127$.}  
\beq
\lambda_{\rm susy} = \frac{g_2^2(\Lambda_{\rm susy})+g_Y^2(\Lambda_{\rm susy})}{8}\,\cos^22\beta.
\eeq{lsusy}
Assuming that all non-SM particles have masses at or above $\Lambda_{\rm susy}$, we use the SM beta functions at two-loop order, and the values of SM couplings at the weak scale given in Ref.~\cite{Buttazzo:2013uya}, to obtain $\lambda_{\rm SM}(\Lambda_{\rm susy})$. We find that accommodating the 125 GeV Higgs in the minimal CCT model, with no additional contributions to the quartic, requires 
\beq
\Lambda_{\rm susy}\gsim 100~{\rm TeV}.
\eeq{Lbig}
This is clearly a much stronger constraint than the experimental bounds considered above, and a model with such a high SUSY-breaking scale would require a very significant amount of fine-tuning: very roughly, fine-tuning can be estimated as $(v/\Lambda_{\rm susy})^2\sim 10^{-6}$. Moreover, for $\tan\beta\approx 1.0$, which is preferred from the point of view of the PEW and direct constraints, a much higher SUSY-breaking scale is required, since $\lambda_{\rm susy}$ is suppressed. 

\begin{figure}[tb]
\begin{center}
\centerline {
\includegraphics[width=4.0in]{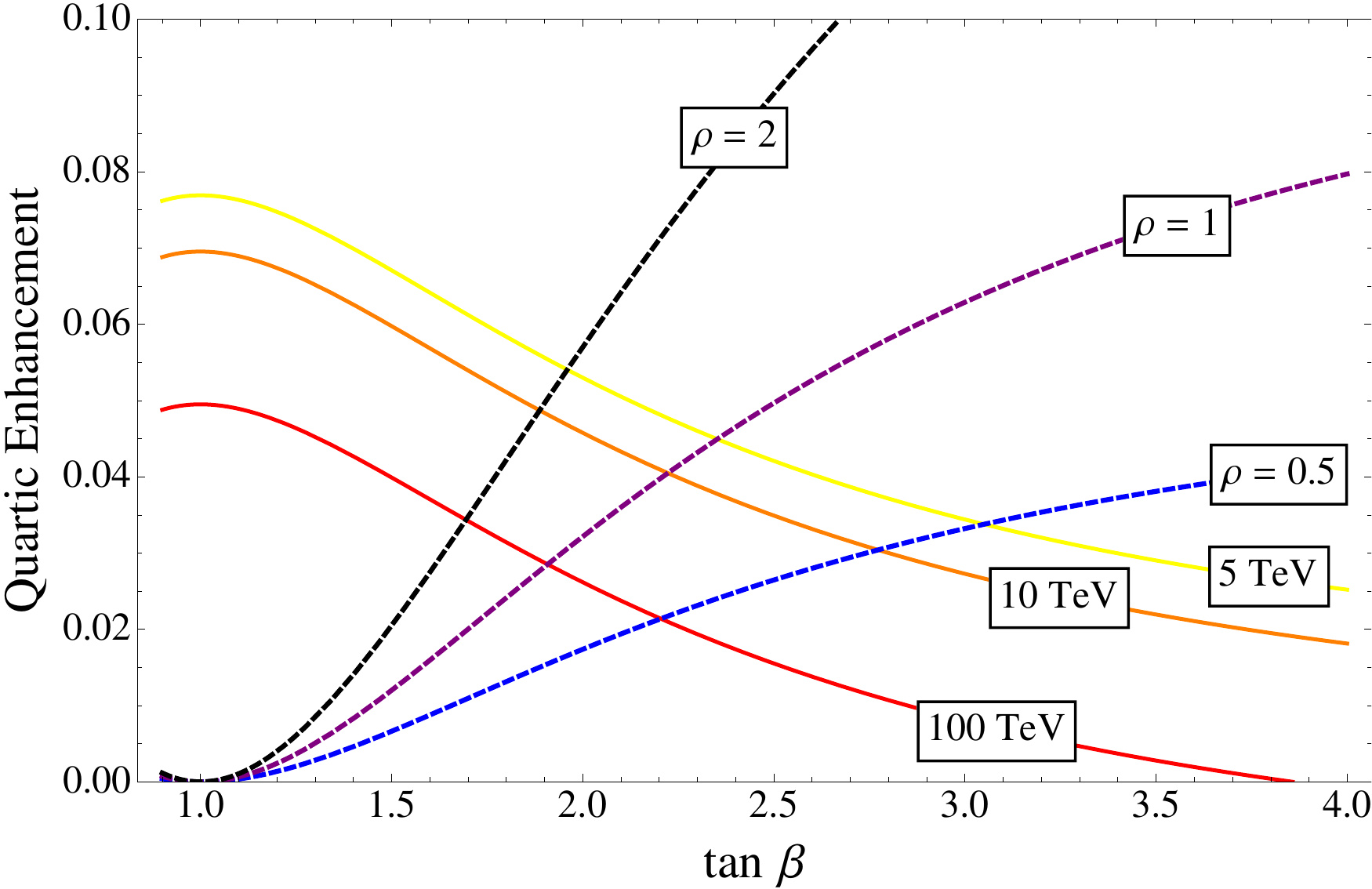}
}
\caption{Solid lines: The difference $\delta$ between the value Higgs quartic $\lambda_{\rm SM}(\Lambda_{\rm susy})$ needed to accommodate the 125 GeV Higgs, and the value predicted by a SUSY theory with the SM gauge group. Top to bottom: $\Lambda_{\rm susy}=5, 10, 100$ TeV. Dashed lines: The additional contribution to $\lambda$ from non-decoupling D-terms possibly present in the CCT model. Top to bottom: $\rho=2.0, 1.0, 0.5$. (For definition of $\rho$ and other details, see Appendix~\ref{app:NDDT}.)}
\label{fig:quartic}
\end{center}
\end{figure}

However, simple extensions of the minimal setup can easily alleviate this tension. For example, consider the scenario in which the gauge symmetry breaking occurs below the SUSY-breaking scale, $f_i< \Lambda_{\rm susy}$. In this case, $\lambda_{\rm susy}$ receives additional contributions from the D-terms associated with non-SM gauge generators, the ``non-decoupling D-terms"~\cite{Batra:2003nj,Maloney:2004rc}. The non-decoupling D-terms in the CCT model were considered in Ref.~\cite{Cai:2012bsa}. They can be obtained as follows.  Introduce additional superfields $A_{2,3}$ (in the adjoint representations of  $SU(2)$ and $SU(3)$, respectively) and $S_{2,3}$ (both singlets under $G$), with a superpotential\footnote{Our model of the non-decoupling D-terms differs slightly from Ref.~\cite{Cai:2012bsa} in that we include soft mass terms in the scalar potential, allowing for a simpler field content and superpotential. For details, see Appendix~\ref{app:NDDT}.}
\beq
W_{\rm new}=  \lambda_{S2}S_2 \Phi_2 \overline{\Phi}_2 + \lambda_{S3}S_3 \Phi_3 \overline{\Phi}_3 + \lambda_{A2}\overline{\Phi}_2 A_2^a \frac{\sigma^a}{2}\Phi_2 + \lambda_{A3}\overline{\Phi}_3 A_3^m G^m\Phi_3. 
\eeq{Wnew}
When the link fields $\Phi$ and $\bar{\Phi}$ acquire vacuum expectation values, F-terms for $S$ are generated, inducing ``hard" F-term SUSY-breaking and prevent the complete decoupling of the ultraviolet D-terms. The UV value of the Higgs quartic is modified as follows:
\beq
\lambda^{\rm NDDT}_{\rm susy} = \frac{\Delta_2 g_2^2(\Lambda_{\rm susy})+\Delta_Y g_Y^2(\Lambda_{\rm susy})}{8}\,\cos^22\beta\,,
\eeq{newlambda} 
where $\Delta_2$ and $\Delta_Y$ are order-one coefficients which can be calculated in terms of the superpotential couplings and soft SUSY-breaking terms. (For details, see Appendix~\ref{app:NDDT}.) In Fig.~\ref{fig:quartic}, we compare the size of the quartic correction required to accommodate a 125 GeV Higgs, defined as $\delta\lambda= \lambda_{\rm SM}(\Lambda_{\rm susy})-\lambda_{\rm susy}$, with the non-decoupling D-term contribution for reasonable model parameters. It is clear that the D-term contribution can easily be large enough to provide a viable model with $\Lambda_{\rm susy}$ in the $5-10$ TeV range. Thus, we conclude that in the presence of non-decoupling D-terms, the 125 GeV Higgs does not place constraints beyond those already known from PEW fits and direct searches. The required fine-tuning is roughly of order $10^{-3}$. The only problematic region is around $\tan\beta=1$, where all D-term contributions to quartic vanish as $\cos^22\beta$. In this region, either a much higher value of $\Lambda_{\rm susy}$, or an alternative mechanism for raising the quartic ({\it e.g.} large threshold corrections), is required.   

Note that the $A$ fields introduced in this section will affect the $\beta$ function coefficients, potentially shifting the location of Landau poles and modifying the constraints on the parameter space in Eq.~\leqn{TanBetaConstraints}. We find that the only effect this has is on the lower bound on $\tan\beta$, which is raised from 0.8 to 0.95. This does not have a significant effect on the precision electroweak and direct constraints on the model.

\section{Higgs Couplings to Photons and Gluons}
\label{sec:HC}

Following the discovery of the Higgs boson, a multi-year program to precisely measure the Higgs couplings is envisioned~\cite{Dawson:2013bba}. The upcoming LHC experiments as well as, hopefully, experiments at a next-generation electron-positron Higgs factory~\cite{Baer:2013cma,Gomez-Ceballos:2013zzn}, will be able to measure many Higgs couplings with precision of $\sim 1$\% or better. It is therefore worthwhile to study deviations from the SM predicted by models of new physics at the TeV scale.

\begin{figure}[tb]
\begin{center}
\centerline {
\includegraphics[width=4.0in]{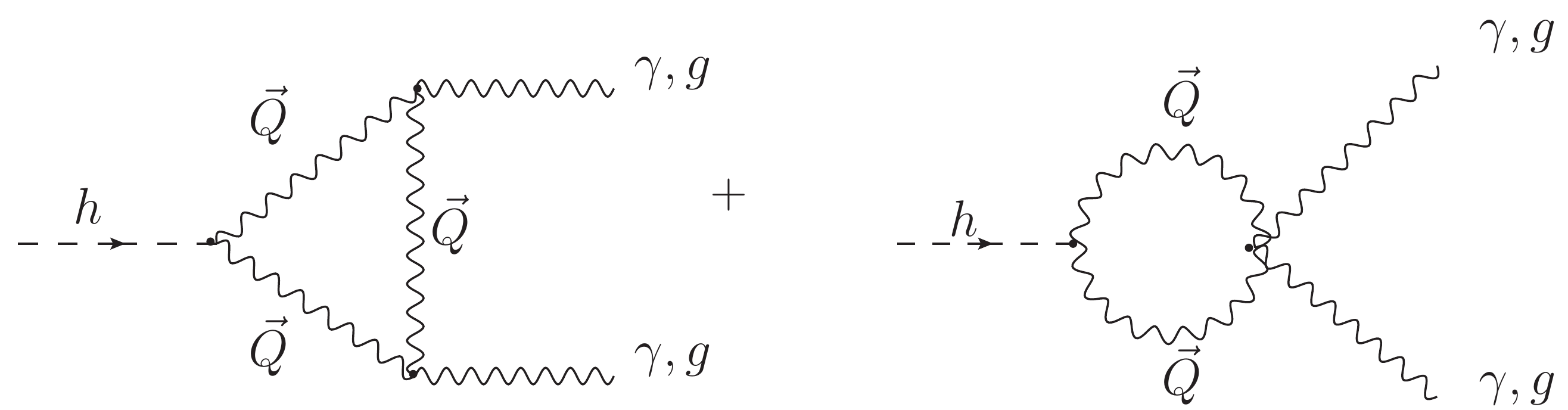}
}
\caption{Swan contribution to Higgs couplings to gluons and photons, at the one-loop level.}
\label{fig:hVV_swans}
\end{center}
\end{figure}

In the CCT model, the corrections to Higgs couplings are of two types. First, since the full structure of the MSSM is reproduced, the Higgs sector is extended to a two-Higgs doublet model, leading to tree-level shifts in the Higgs couplings to gauge bosons and fermions. These effects have been already comprehensively studied in the MSSM~\cite{Djouadi:2005gj}. More interesting are the corrections from new particles running in loops. In particular, it has been argued in Ref.~\cite{Low:2009di,Carmi:2012yp,Carmi:2012in,Farina:2013ssa} that very generally, loops of top quark partners ({\it i.e.}, particles whose loops cancel the quadratic divergence in $m_h^2$ induced by the SM top loop) induce potentially observable shifts in the $hgg$ and $h\gamma\gamma$ couplings.\footnote{These two couplings are singled out because they are absent at tree level in the SM, making the new physics effects relatively more significant. Top partner loops may have other potentially observable effects, {\it e.g.} wavefunction renormalization corrections which may be measured in the $e^+e^-\to hZ$ process at Higgs factories~\cite{Craig:2013xia}.} The corrections from  spin-0 and spin-1/2 top partners have been previously calculated. Here, we focus on the effect of the spin-1 top partner loops, shown in Fig.~\ref{fig:hVV_swans}. We performed the calculation using the {\tt Mathematica} implementation of the $h\to VV$ decay amplitudes for a generic gauge extension of the SM, described in~\cite{Bunk:2013uea} and available on the website~{\tt http://www.phy.syr.edu/~jhubisz/HIGGSDECAYS/}. To leading order in $(m_h/M_{\vec{Q}})^2$, we obtain the effective Lagrangian
\beq
\mathcal{L}_{h\gamma \gamma}= \dfrac{2\alpha}{9 \pi v} C_{\gamma} h F_{\mu \nu} F^{\mu \nu},~~~~\mathcal{L}_{hgg}= \dfrac{\alpha_s}{12 \pi v} C_{g} h G^a_{\mu \nu} G^{a\mu \nu}\,,
\eeq{Leff} 
where $F$ and $G^a$ ($a=1\ldots 8)$ are the SM $U(1)$ and $SU(3)$ field strength tensors, respectively, and the Wilson coefficients are 
\beq
C_g = C_\gamma \,=\, \frac{21}{4} \frac{\hat{g}_5^2 v^2}{m^2_{\vec Q}}\,.
\eeq{Cs}
Here the normalization of $C_g$ and $C_\gamma$ is such that the SM top loop contribution, in the low-$m_h$ limit, is 1. Note that, due to a large numerical coefficient, the swan induces a much larger deviation of the $hgg/h\gamma\gamma$ couplings from their SM values than either a spin-0 stop or a spin-1/2 top partner of the same mass. We find that even very strong bounds on the swan mass discussed above do not completely preclude a potentially observable deviation: a 5 TeV swan, at $\tan\beta=1.0$, induces a fractional shift in the $hgg/h\gamma\gamma$ couplings of about 3\%, which may be within a 3-sigma detection reach at the proposed $e^+e^-$ Higgs factories. 

\begin{figure}[tb]
\begin{center}
\centerline {
\includegraphics[width=3.0in]{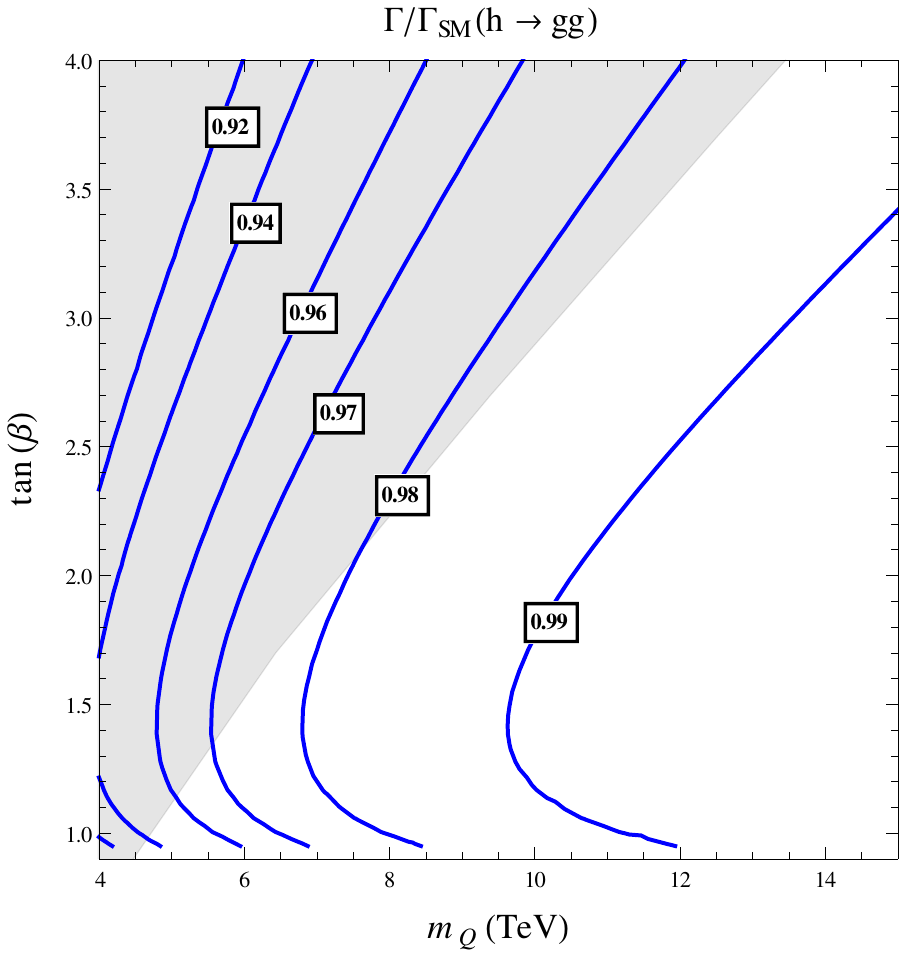}
\includegraphics[width=3.0in]{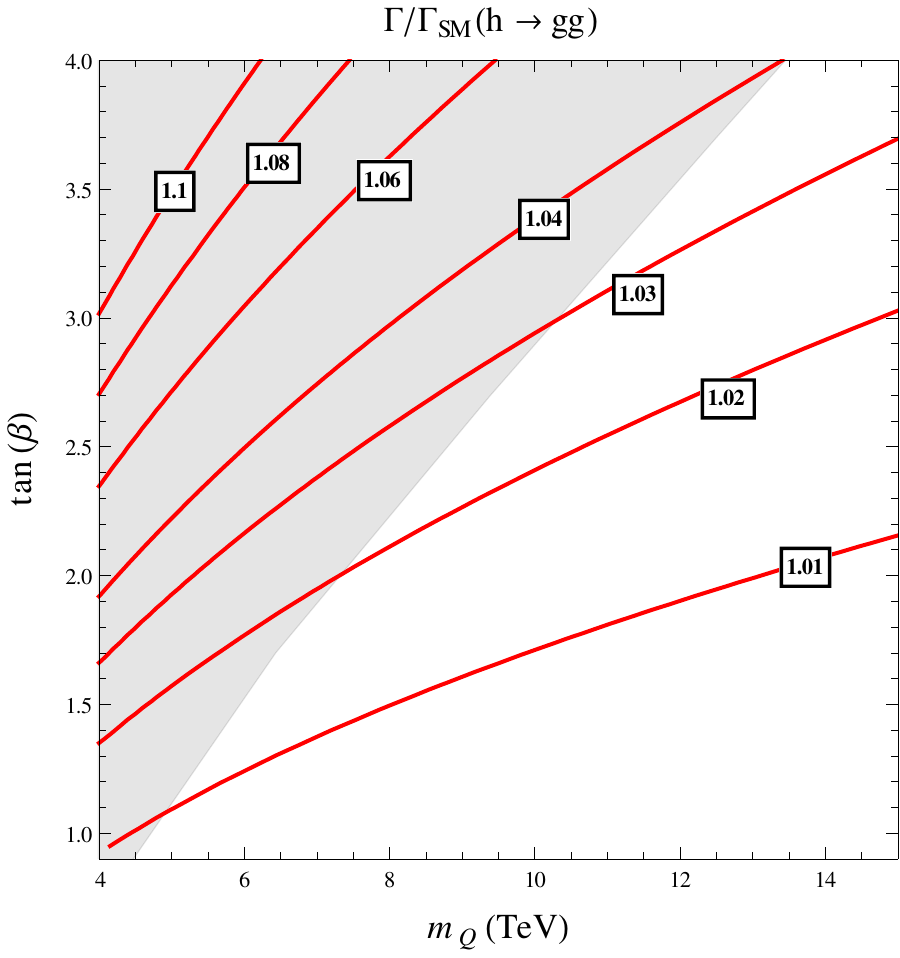}
}
\caption{Fractional deviation in the $hgg$ (left panel) and $h\gamma\gamma$ (right panel) couplings from the SM in the CCT model, as a function of the swan mass and $\tan\beta$. (See text for details on the values of other model parameters.) The shaded region is disfavored by precision electroweak constraints and direct LHC searches for a $Z^\prime$.}
\label{fig:HCs}
\end{center}
\end{figure}

The CCT model contains a large number of colored and/or electrically charged states at the same mass scale as the swan, and loops of those particles will in general contribute to the coefficients $C_g$ and $C_\gamma$, modifying the predictions~\leqn{Cs}. The contributions of scalars and fermions can be computed using the Higgs low energy theorems~\cite{Ellis:1975ap,Shifman:1979eb}, while the spin-1 states other than the swan can be treated using the results of~\cite{Bunk:2013uea}. A comprehensive analysis of these effects is complicated by the large dimensionality of the parameter space. We will not attempt such an analysis here; instead, we illustrate the typical size of the overall contribution to $C_g$ and $C_\gamma$ with a two-dimensional plot, Fig.~\ref{fig:HCs}, where we vary the swan mass and $\tan\beta$ and fix all other parameters. (All parameters with dimension of mass are fixed at the scale $m_{\vec{Q}}$, with mild hierarchies imposed in some cases to ensure that the conditions~\leqn{LambdaTopConditions} are satisfied and an acceptable Higgs mass is generated through non-decoupling D-terms.) In this slice of the parameter space, we find that deviations in the $hgg$ coupling of about 5\% are possible, while the maximum deviation in $h\gamma\gamma$ is about 4\%. Such shifts may be within reach of the proposed $e^+e^-$ Higgs factories. 

\section{Future Prospects for Direct Searches}
\label{sec:100tev}

\begin{figure}[tb]
\begin{center}
\centerline {
\includegraphics[width=4.0in]{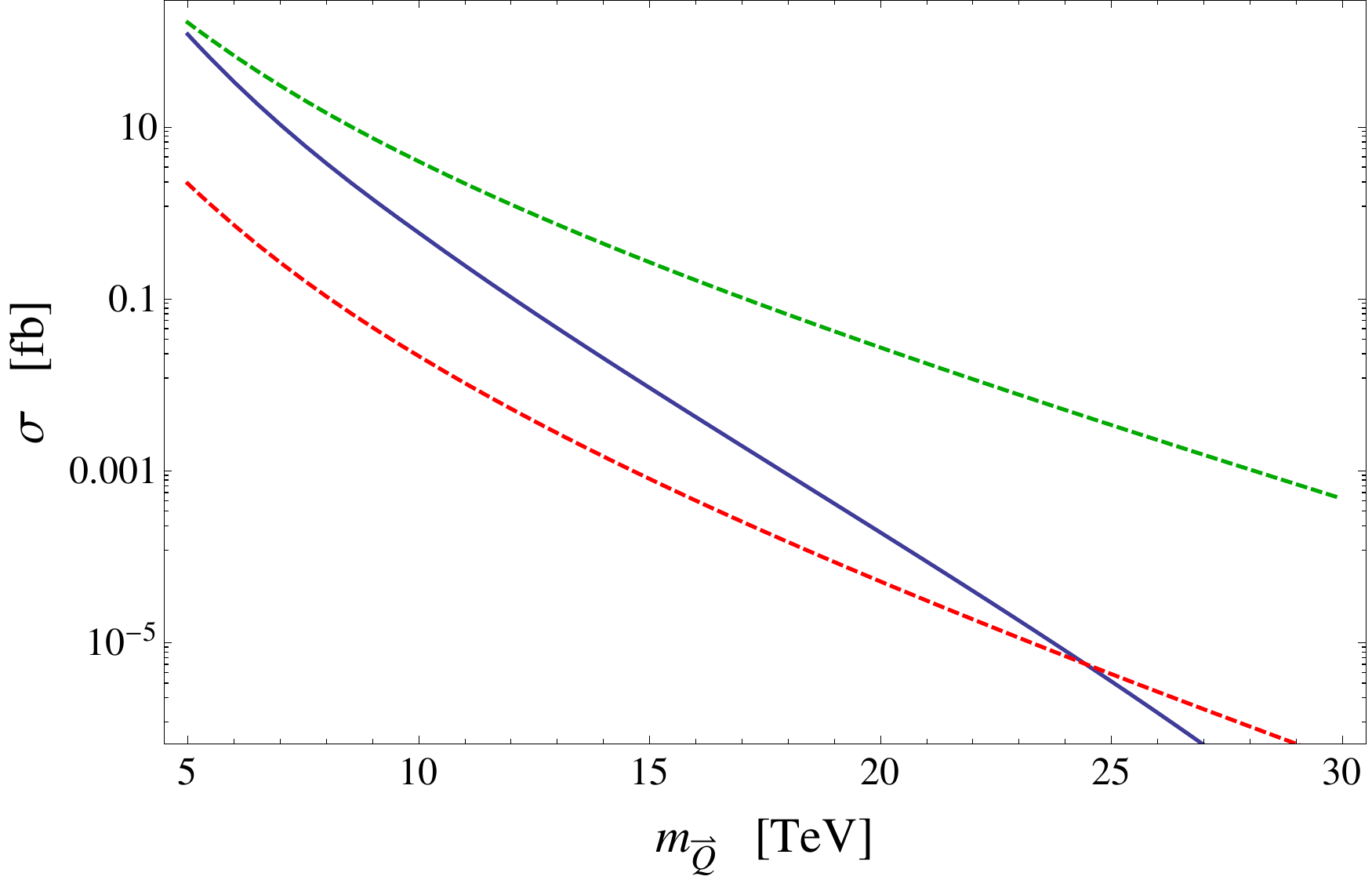}
}
\caption{Swan production cross sections at a 100 TeV $pp$ collider: $pp\to\vec{Q}\vec{Q}$ (blue), $\vec{Q}\tilde{g}$ (green dashed), $\vec{Q}\tilde{\chi}^0_1$ (red dashed).}
\label{fig:100}
\end{center}
\end{figure}

Existing bounds on the swan mass, and the fact that swans must be pair-produced, preclude the possibility of direct swan production at the LHC. Of course, it may well happen that other particles in the CCT model, such as a $Z^\prime$ or some of the MSSM-like states, will be within the reach of the LHC-14. However, without a direct observation of the swan, it would be difficult to distinguish between this model and more conventional realizations of weak-scale supersymmetry. If a $Z^\prime$ is discovered, some indirect evidence can perhaps be obtained by measuring its couplings, which are predicted in the CCT model with few free parameters (see Appendix A). A much more direct and convincing test would have to await the direct discovery of the swan, and measurement of its spin. A next-generation $pp$ collider with $\sqrt{s}=100$ TeV, which is currently under discussion in the high-energy physics community, would provide an opportunity for such a direct discovery. As a first step to an estimate of the potential of such a collider to search for swans, we computed the cross sections of swan pair-production, along with associated production with a gluino $\tilde{g}$ and a neutralino $\tilde{\chi}^0_1$. The analytic formulas for parton-level cross sections are collected in Appendix C. The cross sections for 100 TeV $pp$ collisions are plotted in Fig.~\ref{fig:100}. Here we assumed $m_{\tilde{g}}=1$ TeV and $m_{\tilde{\chi}^0_1}=0.5$ TeV; the plotted associated production cross sections represent the maximum possible values, and would decrease if $m_{\tilde{g}}$/$m_{\tilde{\chi}^0_1}$ are increased. We used the {\tt NNPDF2.3} NNLO parton distribution functions~\cite{Ball:2013hta}, including top quark pdf's for associated production, and set the renormalization/factorization scale to $Q^2=(10~{\rm TeV})^2$. It is interesting to note that the large associated production cross sections are due to appreciable $b$ and $t$ content in the proton at this scale. 

Swans within a broad mass range will be copiously produced in 100 TeV $pp$ collisions. For example, if 3000 fb$^{-1}$ of data is collected (the integrated luminosity assumed in the Snowmass study~\cite{Avetisyan:2013onh}), we expect that $\gsim {\cal O}(100)$ swans would be produced in pair-production up to $m_{\vec{Q}}\approx 15$ TeV, and in association with gluinos up to $m_{\vec{Q}}\approx 25$ TeV (assuming $m_{\tilde{g}}\ll m_{\vec{Q}}$). This suggests that direct reach of such an experiment for swan discovery can potentially extend into $10-20$ TeV domain, although the actual reach depends on the swan decay chains, which will determine relevant backgrounds, kinematic cuts, {\it etc.} Once a swan is produced, its spin could be determined using the techniques proposed for top partner spin determination at the LHC, see {\rm e.g.}~\cite{Chen:2012uw}. Thus, a 100 TeV collider may be capable of directly demonstrating the existence of a spin-1 top partner.

\section{Conclusions and Outlook}
\label{sec:conc}

In this paper, we considered the phenomenology of the Cai-Cheng-Terning (CCT) model, in which the superpartner of the (left-handed) SM top quark is the spin-1 particle, the ``swan". Our main result is that existing constraints from precision electroweak fits and direct LHC searches for a $Z^\prime$ place a very strong bound on the swan mass, which is required to be above at least 4.5 TeV, and in fact above 10 TeV in most of the parameter space. The primary reason for this bound is the tight relation between the swan mass and the mass of a neutral, $R$-even $Z^\prime$ boson, which is tightly constrained. The masses of the two bosons arise from the same symmetry breaking, and the structure of the gauge couplings of the CCT model induces an additional hierarchy, typically of a factor $5-10$, between the swan and $Z^\prime$ masses. 

The tight bounds on the swan mass imply that the models of this type would need to be quite fine-tuned if realized in nature, making them less appealing. It also precludes the possibility of a direct swan discovery at the LHC. It is interesting to note, however, that neither conclusion would hold in a model with a spin-1 top partner {\it not} accompanied a $Z^\prime$ whose mass arises from the same symmetry breaking, or in a model where a $Z^\prime$ is odd under an $R$ parity. It would be interesting to construct such models. Even if a complete model proves hard to build, a phenomenological model with these features, analogous to minimal set-ups used for the spin-0 top partner (``natural SUSY"~\cite{Brust:2011tb,Papucci:2011wy}) and the spin-1/2 top partner (see {\it e.g.}~\cite{Berger:2012ec}), would be potentially quite useful.

\vskip0.8cm
\noindent{\large \bf Acknowledgments} 
\vskip0.3cm

We would like to thank Christophe Grojean, Jay Hubisz, and especially Haiying Cai, for useful discussions. J.C., M.P. and N.R.L. are supported by the U.S. National Science Foundation through grant PHY-0757868. M.P. is also supported by CAREER grant PHY-0844667. B.J. thanks Cornell University for hospitality throughout the course of this work. B.J. is supported by the Department Of Energy under grant DE-FG02-85ER40237.

\begin{appendix}

\section{Appendix A: Masses and Couplings of $Z^\prime$ States}
\label{app:Zprime}

Compared to the MSSM, this model possesses three additional neutral, massive gauge bosons. Two of them are linear combinations of the UV gauge fields $B_H$, $B_V$, $B_{24}$ obtained by diagonalizing the following quadratic terms: 
\beqa
6\tilde{f}_3^2 \left(\frac{\hat{g}_H}{6}B_H - \frac{\hat{g}_V}{10}B_V - \frac{\hat{g}_5}{\sqrt{15}}B_{24}\right)^2 + 4\tilde{f}_2^2 \left(\frac{\hat{g}_V}{10}B_V - \frac{\sqrt{15}}{10}\hat{g}_5B_{24}\right)^2  ~.
\eeqa{QuadBTerms}
The massless linear combination $B \equiv \frac{g_Y}{\hat{g}_H}B_H + \frac{g_Y}{\hat{g}_V}B_V + \frac{g_Y}{\sqrt{15}\hat{g}_5}B_{24}$ will be the gauge boson of the SM $U(1)_Y$ group; we refer to the other two eigenstates with non-vanishing masses as the $Z'$ and $Z''$, in ascending order of masses. As discussed in section \ref{sec:model}, it is convenient to re-express the parameters $\hat{g}_H$, $\hat{g}_V$, $\hat{g}_5$, $\tilde{f}_2$, and $\tilde{f}_3$ in terms of $\epsilon \equiv g_Y^2/15\hat{g}_5^2$, $\theta \equiv \arctan \left(\hat{g}_V/\hat{g}_H\right)$, $\tilde{f}^2 \equiv \tilde{f}_2^2 + \tilde{f}_3^2$ and $\phi \equiv \arctan \left(\tilde{f}_2/\tilde{f}_3\right)$. In this parameterization, the mass of the $Z'$ and $Z''$ can be written as:
\beqa
m^2_{Z',Z''} = \frac{m^2_{\vec{Q}}}{20\left(1-\epsilon\right)} \left(A(\epsilon,\theta,\phi) \mp \sqrt{B(\epsilon,\theta,\phi)}\right)~,
\eeqa{NewZsFullMass}
where $m^2_{\vec{Q}}$ is the squared mass of the swan and the $A$, $B$ functions are given by:
\beqa
A(\epsilon,\theta,\phi) \equiv 50 \epsilon \csc^2{\theta}\cos^2{\phi} + 3 \epsilon \sec^2{\theta}\left(\cos{2\phi}+5\right)-2\left(1-\epsilon\right)\left(\cos{2\phi}-5\right)~,
\eeqa{NastyAFunction}
and,
\beqa
B(\epsilon,\theta,\phi) &\equiv& 2500\epsilon^2 \csc^4{\theta}\cos^4{\phi}+9 \epsilon^2 \sec^4{\theta}\left(\cos{2\phi}+5\right)^2  \nn \\
&+& 100 \epsilon \csc^2{\theta}\cos^2{\phi}\left(5\left(\epsilon+2\right)\cos{2\phi} + 5\epsilon -2 \right) \nn \\
&+& 3\epsilon \sec^2{\theta}\left(300 \epsilon \cos{2\phi} + \left(27 \epsilon + 98\right)\cos{4\phi} + 177 \epsilon -2 \right)~.
\eeqa{NastyBFunction}
Since $\epsilon$ is typically $O(5) \times 10^{-3}$ (see Eq. (\ref{epsdef})), we can obtain much simpler formulas by expanding to $O(\epsilon)$, in which case we can write the $Z'$ mass as:
\beq
m_{Z^\prime}^2 \approx 30 m_{\vec{Q}}^2 \epsilon \frac{\csc^2{2\theta} \sin^2{2 \phi}}{5-\cos{2\phi}} = 2 g_Y^2 \frac{\csc^2 2\theta \sin^2 2\phi}{5 - \cos 2\phi}\,\tilde{f}^2,
\eeq{Zpmassapprox2}
where the second equality was obtained by using $m_{\vec{Q}}^2 = \hat{g}_5^2 \tilde{f}^2$ and the definition of $\epsilon$. For the $Z''$, we have: 
\beq
m_{Z''}^2 \approx m_{\vec{Q}}^2 \left(\frac{5-\cos{2\phi}}{5}\right) + O(\epsilon)~.
\eeq{Zppmassapprox2}
The couplings of the $Z'$ to the light fermions of the SM will be given by
\beq
g_{Z' \bar{f} f} = \hat{g}_H \langle Z'|B_H \rangle \left(Q-T_3\right)~,
\eeq{lightfermcoup}
where $\langle Z'|B_H \rangle$ is the amount of $B_H$ contained in the $Z'$ mass eigenstates. The couplings of the $Z''$ follows an analogous formula, with the replacement of $\langle Z'|B_H \rangle$ by $\langle Z''|B_H \rangle$. While explicit formulas for these coefficients are straightforward to compute, they are cumbersome and unenlightening. We note, however, that $\hat{g}_H  \langle Z'|B_H \rangle= \dfrac{| g_Y\tan^{-1}{\theta}|}{\sqrt{15}} + O(\epsilon)$, which indicates that the $Z'$ decouples from the light SM fermions in the large $\tan{\theta}$ region; this explains why the bounds on the $Z'$ mass are weakest in this region of Fig.~\ref{fig:Direct}. The couplings of the $Z'$ and $Z''$ to the third generation quarks will be different from Eq. (\ref{lightfermcoup}) because these fermions are charged differently under the UV gauge group\footnote{The exception is the right-handed bottom quark $b_R$, whose coupling to the $Z'$ follows Eq. (\ref{lightfermcoup})}. The coupling to the right-handed top is
\beq
g_{Z'  \bar{t}_R t_R} = \frac{1}{2} \hat{g}_H \langle Z'|B_H \rangle - \frac{1}{10}\hat{g}_V \langle Z'|B_V \rangle - \frac{1}{\sqrt{15}}\hat{g}_5 \langle Z'|B_{24} \rangle~,
\eeq{righthandtopcoup}
while the coupling to the third generation doublet of the SM, $Q_L^3 = (t_L, b_L)$ is
\beq
g_{Z'  \bar{Q}_L^3 Q_L^3} = \sqrt{\frac{5}{12}}\hat{g}_5 \langle Z'|B_{24} \rangle~.
\eeq{thirdgendoubletcoup}
The couplings of the $Z''$ can once again be obtained by replacing $\langle Z'|B_i \rangle$ by $\langle Z''|B_i \rangle$ in the above.


\section{Appendix B: Non-Decoupling D-Terms}
\label{app:NDDT}

The non-decoupling D-terms coefficients $\Delta_2$ and $\Delta_Y$ were introduced in Section~\ref{sec:HiggsMass} as a way of enhancing the tree-level quartic of the Higgs at the scale $\Lambda_{\rm susy}$ to obtain the observed Higgs mass. (Non-decoupling D-terms in the CCT model were previously discussed in Ref.~\cite{Cai:2012bsa}.) Here we outline the derivation of these coefficients. 

Combining the superpotential terms of Eqs. \ref{SuperPot} and \ref{Wnew} to the usual soft SUSY breaking terms, we obtain the following potential for the link fields:
\beqa
V_{link} &=& \left(\mu_2^2 + m_2^2\right)\Phi_2 \Phi_2^{*} + \left(\mu_2^2 + \overline{m}_2^2\right)\overline{\Phi}_2 \overline{\Phi}_2^{*} + \left(\mu_3^2 + m_3^2\right)\Phi_3 \Phi_3^{*} + \left(\mu_3^2 + \overline{m}_3^2\right)\overline{\Phi}_3 \overline{\Phi}_3^{*} \nn \\ &-& b_2 \left( \Phi_2 \overline{\Phi}_2 + c.c.\right)- b_3 \left(\Phi_3 \overline{\Phi}_3 + c.c.\right) + y_1^2 |\Phi_3 \overline{\Phi}_2|^2 + \lambda_{S2}^2 |\Phi_2 \overline{\Phi}_2|^2 + \lambda_{S3}^2 |\Phi_3 \overline{\Phi}_3|^2 \nn \\ &+& \lambda_{A2}^2|\overline{\Phi}_2 ^a \frac{\sigma^a}{2}\Phi_2|^2 + \lambda_{A3}^2 |\overline{\Phi}_3 ^m G^m\Phi_3|^2 + \left({\rm D-terms}\right)~.
\eeqa{LinkPot}
Though the soft SUSY-breaking masses $m^2_i$ and $\overline{m}^2_i$ can in principle be independent from one another, we will make the simplifying assumption that they are identical. Note however that while this assumption greatly simplifies the following analysis, the theory possesses no symmetry that could make this equality exact and stable under radiative corrections, even if it is approximately realized at the messenger scale. Under this assumption then, we can derive simple formulas for the vevs from Eq. (\ref{LinkPot}):
\beqa
f_2^2 = \overline{f}_2^2 = \frac{b_2 - \left(\mu_2^2 + m_2^2\right)}{2\lambda_{S2}^2}~, \nn \\
f_3^2 = \overline{f}_3^2 = \frac{b_3 - \left(\mu_3^2 + m_3^2\right)}{3\lambda_{S3}^2}~.
\eeqa{VEVFormula}
We can shift the link fields by these vevs in Eq. (\ref{LinkPot}) and compute the mass spectrum for the scalar components of the link sector. It is convenient to invert the formulas for the masses to express the parameters of the potential in terms of more physical quantities: the vevs $f_2$ and $f_3$, the masses of the two CP-odd singlets $m^2_{O_{2,3}}$, and the masses of the two CP-even singlets $m^2_{E_{2,3}}$. The relationship between the masses and the parameters of the potential in Eq.~(\ref{LinkPot}) is:
\beqa
m^2_{O_{2,3}} = 2 b_{2,3} ~,\\
m^2_{E_2} = 4 f_2^2 \lambda^2_{S2} ~,\\
m^2_{E_3} = 6 f_3^2 \lambda^2_{S3} ~.
\eeqa{PhysicalPar}
The effect of the aforementioned non-decoupling D-terms on the low-energy Higgs potential can be obtained by integrating out at tree-level the scalar fields that possess trilinear coupling to the Higgs bilinears. This will effectively modify the low-energy Higgs potential through the substitutions $g_Y \rightarrow \Delta_Y g_Y$, $g_2 \rightarrow \Delta_2 g_2$, where:
\beqa
&\Delta_2& = \left(1+\dfrac{\rho_2}{2\hat{g}_2^2}\right) \times \left(1+\dfrac{\rho_2}{2\left(\hat{g}_5^2+\hat{g}_2^2\right)}\right)^{-1}~, \nn \\
&\Delta_Y& = \dfrac{1 +  N_2\rho_2 +  N_3\rho_3 + N_{23}\rho_2 \rho_3}{1+ D_2\rho_2 + D_3\rho_3 + D_{23}\rho_2 \rho_3}~,
\eeqa{NDDTEnhancement}
\normalsize
with 
\beqa
\rho_2 \equiv \frac{m_{O2}^2-m_{E2}^2}{f_2^2} = 2\left(\dfrac{m_2^2 + \mu_2^2}{f_2^2}\right), \, \, \, \, \, \, \, \, \, \, \, \, \, \, \, \, \rho_3 \equiv \frac{m_{O3}^2-m_{E3}^2}{f_3^2}= 2\left(\dfrac{m_3^2 + \mu_3^2}{f_3^2}\right)~,
\eeqa{rhodef}
and the various $N_i(\theta, \epsilon)$, $D_i(\theta, \epsilon)$ coefficient functions are:
\beqa
N_2(\theta, \epsilon) &\equiv& \left(\dfrac{1+15\epsilon}{2g_Y^2}\right)~, \nn \\
N_3(\theta, \epsilon) &\equiv& 3\left(\dfrac{\epsilon\sin^2{\theta} +\cos^2{\theta}}{g_Y^2}\right)~, \nn \\
N_{23} (\theta, \epsilon) &\equiv& 3\left(\dfrac{\left(1-\epsilon\right)\sin^2{\theta}\cos^2{\theta}\left(1 +\epsilon \tan^2{\theta} + 25 \epsilon \csc^2{\theta} \right)}{2g_Y^4} \right)~, \nn \\
D_2(\theta, \epsilon) &\equiv& \left(\dfrac{\left(1-\epsilon\right)\left(1+ 33 \epsilon + 16 \epsilon \cos{2\theta} -\left( 1- \epsilon \right)\cos{4\theta}\right)}{4g_Y^2}\right)~, \nn \\
D_3(\theta, \epsilon) &\equiv& \left(\dfrac{3\left(1-\epsilon\right)\sin^2{2\theta}\left(1+\epsilon \tan^2{\theta} \right)}{4g_Y^2}\right)~, \nn \\
D_{23}(\theta, \epsilon) &\equiv& \left(\dfrac{75\left(1-\epsilon \right)^2 \epsilon \sin^2{2\theta}}{8g_Y^4}\right)~.
\eeqa{coeffunctions}

\section{Appendix C: Parton-Level Cross Sections for Swan Production}
\label{app:xsec}

In this Appendix, we list the formulas for parton-level cross sections of swan production in $pp$ collisions. For swan pair-production, we find
\beqa
\frac{d \sigma(g g \rightarrow \vec{Q} \bar{\vec{Q}})}{d \cos(\theta)} &=&  \frac{g_3^4}{16 \pi  s} \sqrt{1-\frac{4 m_{\vec{Q}}^2}{s}}  \Bigg[4+ \frac{9 \left(m_{\vec{Q}}^4+m_{\vec{Q}}^2 s-t u\right)}{4 s^2} + \frac{6 m_{\vec{Q}}^4+2 s^2}{3 \left(t-m_{\vec{Q}}^2\right)^2} \nn \\
 & & \hskip-2cm +\frac{6 m_{\vec{Q}}^4+2 s^2}{3 \left(u-m_{\vec{Q}}^2\right)^2} -\frac{\left(m_{\vec{Q}}^2+s\right) \left(m_{\vec{Q}}^2+3 s\right)}{2 s \left(m_{\vec{Q}}^2-u\right)} -\frac{\left(m_{\vec{Q}}^2+s\right) \left(m_{\vec{Q}}^2+3 s\right)}{2 s   \left(t-m_{\vec{Q}}^2\right)}  \Bigg].
\eeqa{XS_GGSS}
The quark-initiated contribution to swan pair-production is negligibly small in the relevant swan mass range. The associated swan-gluino production cross section is
\beqa
\frac{d \sigma(g t_L \rightarrow \vec{Q} \tilde{G})}{d \cos(\theta)} &=&  \frac{g_3^2 \hat{g}_5^2 \cos(\theta_{\tilde{G}})^2 }{16 \pi  s^2} \sqrt{\left(s-m_{\tilde{G}}^2-m_{\vec{Q}}^2\right)^2-4 m_{\tilde{G}}^2 m_{\vec{Q}}^2}
\Bigg[\frac{4 m_{\vec{Q}}^4-m_{\tilde{G}}^4-u \left(m_{\tilde{G}}^2+2 m_{\vec{Q}}^2\right)}{9m_{\vec{Q}}^2 s} \nn \\
 &+& \frac{4 s^2 +4 m_{\vec{Q}}^4 - 2 m_{\tilde{G}}^4 - 2 m_{\tilde{G}}^2 m_{\vec{Q}}^2 }{9 \left(t-m_{\vec{Q}}^2\right)^2}
+\frac{2 m_{\tilde{G}}^2 m_{\vec{Q}}^4-m_{\tilde{G}}^6-m_{\tilde{G}}^4 m_{\vec{Q}}^2}{2   m_{\vec{Q}}^2 \left(u-m_{\tilde{G}}^2\right)^2}  -\frac{1}{18} - \frac{m_{\tilde{G}}^2}{4 m_{\vec{Q}}^2} \nn \\
 &-& \frac{2   m_{\vec{Q}}^2 s^2 -4 s \left(2 m_{\vec{Q}}^4 -  m_{\tilde{G}}^4 - m_{\tilde{G}}^2 m_{\vec{Q}}^2 \right) - 4 m_{\tilde{G}}^6-9 m_{\tilde{G}}^4 m_{\vec{Q}}^2+3 m_{\tilde{G}}^2  m_{\vec{Q}}^4+10 m_{\vec{Q}}^6}
{18 m_{\vec{Q}}^2 s \left(t-m_{\vec{Q}}^2\right)} \nn \\
 &+& \frac{\left(m_{\tilde{G}}^2+2 m_{\vec{Q}}^2\right) \left(s^2-2 s
   \left(m_{\vec{Q}}^2-m_{\tilde{G}}^2\right)-2 m_{\tilde{G}}^2 m_{\vec{Q}}^2+2
   m_{\vec{Q}}^4\right)}{4 m_{\vec{Q}}^2 s \left(m_{\tilde{G}}^2-u\right)}
  \Bigg]\,,
\eeqa{XS_GtSG}
where $\cos(\theta_{\tilde{G}})$ is the overlap of the gaugino being produced with the $SU(5)$ gaugino. (In Fig.~\ref{fig:100}, we assumed that the mixing angle for gauginos and corresponding gauge bosons are aligned.) Finally, the associated swan-neutralino production cross section is
\beqa
\frac{d \sigma(g t_L \rightarrow \vec{Q} \tilde{N})}{d \cos(\theta)} &=&  \frac{g_3^2 \hat{g}_5^2 \cos(\theta_{\tilde{N}})^2 }{16 \pi  s^2} \sqrt{\left(s-m_{\tilde{N}}^2-m_{\vec{Q}}^2\right)^2-4 m_{\tilde{N}}^2 m_{\vec{Q}}^2}
\Bigg[ \frac{1}{4} + \frac{m_{\tilde{N}}^2}{24 m_{\vec{Q}}^2} +    \nn \\
 &+& \frac{t \left(m_{\tilde{N}}^2+2 m_{\vec{Q}}^2\right)-3 m_{\tilde{N}}^2   m_{\vec{Q}}^2+2 m_{\vec{Q}}^4 -2 m_{\tilde{N}}^4}{24 m_{\vec{Q}}^2 s} + \frac{2 s^2+2 m_{\vec{Q}}^4-m_{\tilde{N}}^4-m_{\tilde{N}}^2 m_{\vec{Q}}^2}{12
   \left(t-m_{\vec{Q}}^2\right)^2} \nn \\
 &+& \frac{4 m_{\vec{Q}}^2 s^2 +s \left(2 m_{\vec{Q}}^4  -m_{\tilde{N}}^4  -m_{\tilde{N}}^2 m_{\vec{Q}}^2 \right) -3 m_{\tilde{N}}^2 m_{\vec{Q}}^4+2 m_{\vec{Q}}^6+ m_{\tilde{N}}^6}{12  m_{\vec{Q}}^2 s \left(t-m_{\vec{Q}}^2\right)}
  \Bigg].
\eeqa{XS_GtSN}

\end{appendix}

\bibliography{lit}
\bibliographystyle{jhep}
\end{document}